\documentclass[sigconf]{acmart}

\AtBeginDocument{%
  \providecommand\BibTeX{{%
    \normalfont B\kern-0.5em{\scshape i\kern-0.25em b}\kern-0.8em\TeX}}}

\acmConference[Conference acronym 'XX]{Make sure to enter the correct
  conference title from your rights confirmation email}{June 03--05,
  2018}{Woodstock, NY}

\acmBooktitle{Woodstock '18: ACM Symposium on Neural Gaze Detection,
 June 03--05, 2018, Woodstock, NY} 
\acmISBN{978-1-4503-XXXX-X/18/06}

\usepackage{algorithm}
\usepackage{algpseudocode}
\usepackage{xspace}
\usepackage{multirow}
\usepackage{tipa}
\usepackage{enumitem}
\usepackage{booktabs}
\usepackage{colortbl}
\usepackage{xcolor}
\usepackage{tablefootnote}
\usepackage{tipa}

\begin{document}

\title{Extending Visual Dynamics for Video-to-Music Generation}

\author{Xiaohao Liu}
\email{xiaohao.liu@u.nus.edu}
\affiliation{%
  \institution{National University of Singapore}
  \city{Singapore}
  \country{Singapore}
}

\author{Teng Tu}
\email{teng.tu@u.nus.edu}
\affiliation{%
  \institution{National University of Singapore}
  \city{Singapore}
  \country{Singapore}
}

\author{Yunshan Ma}
\email{ysma@smu.edu.sg}
\affiliation{%
  \institution{Singapore Management University}
  \city{Singapore}
  \country{Singapore}
}
\authornote{Corresponding author.}
\author{Tat-Seng Chua}
\email{dcscts@nus.edu.sg}
\affiliation{%
  \institution{National University of Singapore}
  \city{Singapore}
  \country{Singapore}
}

\begin{abstract}

Music profoundly enhances video production by improving quality, engagement, and emotional resonance, sparking growing interest in video-to-music generation. 
Despite recent advances, existing approaches remain limited in specific scenarios or undervalue the visual dynamics. 
To address these limitations, we focus on tackling the complexity of dynamics and resolving temporal misalignment between video and music representations. 
To this end, we propose DyViM, a novel framework to enhance dynamics modeling for video-to-music generation.
Specifically, we extract frame-wise dynamics features via a simplified motion encoder inherited from optical flow methods, followed by a self-attention module for aggregation within frames. 
These dynamic features are then incorporated to extend existing music tokens for temporal alignment.
Additionally,  high-level semantics are conveyed through a cross-attention mechanism, and an annealing tuning strategy benefits to fine-tune well-trained music decoders efficiently, therefore facilitating seamless adaptation.
Extensive experiments demonstrate DyViM's superiority over state-of-the-art (SOTA) methods.

\end{abstract}



\definecolor{dynamic}{HTML}{ef897f}
\definecolor{semantic}{HTML}{7eab55}
\definecolor{music}{HTML}{b89230}

\newcommand{\cdynamic}[1]{{\color{dynamic}{#1}}}
\newcommand{\csemantic}[1]{{\color{semantic}{#1}}}
\newcommand{\cmusic}[1]{{\color{music}{#1}}}

\newcommand{\eg}{\emph{e.g.}\xspace}
\newcommand{\ie}{\emph{i.e.}\xspace}

\maketitle

\section{Introduction}

Video-to-music generation, the task of creating music that temporally aligns with visual content, has garnered growing interest due to its potential to enrich user experience in media consumption~\cite{Survey-Deep-Music-Generation, survey-music-gen, V2Meow}.
This task requires capturing not only the visual semantics (\eg, general mood or themes of a video) but also fine-grained dynamics (\eg, camera movements or scene transitions) in real-time, ensuring that the music reflects nuanced shifts within the video flow~\cite{CMT, Video2Music}. 
Notably, visual dynamics play a substantial role in establishing video-music correlations~\cite{Diff-BGM}. 
For instance, videos may share similar semantics, yet their visual dynamics differ, making them an indispensable factor for music cues~\cite{D2MGAN, CDCD, LORIS, VMAS}.

\begin{figure}
    \centering
    \includegraphics[width=0.99\linewidth]{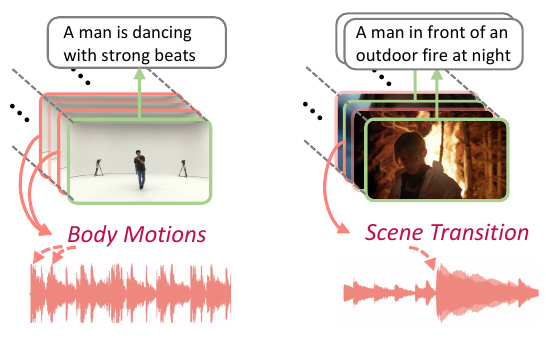}
    \caption{Illustrative examples of video-music pairs of dance video (left) and music video (right). 
    Visual {\color{dynamic}{dynamics}} (\eg, body motions or scene transitions) demonstrates temporal synchronization with music rhythm and changes (\eg, beats).  
    }
    \label{fig:top_figure}
\end{figure}

Unfortunately, existing methods in video-to-music generation exhibit limitations from the perspective of visual dynamics. We group them into three main approaches.
One approach leverages human-centric dynamics, like body motions~\cite{D2MGAN, CDCD, LORIS} or gestures~\cite{FoleyMusic, Sight2Sound}, which work limited in specific scenarios like dance or instrumental videos. 
Another approach maps quantitative visual dynamics, such as motion magnitude or speed, to musical concepts, like beats and note density, with rule-based methods~\cite{CMT, Video2Music}, while this relies heavily on expert knowledge. 
A third approach utilizes various dynamics features extracted from pre-trained models~\cite{V2Meow}, such as I3D~\cite{I3D}; however, it struggles to achieve precise temporal synchronization.
Overall, despite dynamics being involved, these methods remain limited by their inability to generalize nuanced dynamics features and to finely control music generation in sync with temporal changes. 
These unresolved limitations hinder the generation of music that dynamically adapts to diverse visual cues.

Addressing the challenges in video-to-music generation requires tackling two core issues: effectively capturing complex visual dynamics and resolving the representational misalignment between video and music.
First, visual dynamics are intricate and multifaceted, serving as essential cues for musical changes. 
For instance, as shown in Figure~\ref{fig:top_figure}, visual dynamics can manifest as body movements in dance videos, directly indicating musical beats, or as scene transitions and camera movements in music videos (MVs) that enhance storytelling. 
Music aligns with these nuanced visual dynamics, rather than merely reflecting high-level semantics, making it more distinct and contextually relevant to each video. 
However, recent approaches to handle visual dynamics tend to be superficial, often combining various video encoders without adapting them specifically for music generation, limiting both the generalization for various scenarios and effectiveness~\cite{V2Meow, LORIS}.
Second, video and music possess inconsistency in their representations, posing challenges for achieving fine-grained temporal synchronization. 
Video typically operates at frame rates of around 24 frames per second, whereas music is sampled at a much higher rate, such as 32 kHz.
Despite sharing the same duration in a video-music pair, these differences in data density and structure complicate alignment.  
This inconsistency leads recent methods to rely on coarse-level (\eg, short-term context~\cite{Diff-BGM}) alignment, resulting in suboptimal synchronization. 
Although some video-to-audio methods~\cite{FoleyCraft, CondFoelyGen} suggest using onset detection to enforce synchronization, they sacrifice flexibility for music generation (\ie, not every dynamics necessarily indicates a musical change and vice versa).
Encoding informative visual dynamics flexibly, along with its fine-grained conditioning for music generation across diverse scenarios, remains largely unexplored.

To this end, we introduce a novel framework for enhancing \underline{\textbf{Dy}}namics modeling for \underline{\textbf{Vi}}deo-to-\underline{\textbf{M}}usic generation, named as \textbf{DyViM} (\textipa{/""daI.vIm/}). 
At its core, we propose to model visual dynamics by 1) encoding nuanced frame-wise dynamics features using an optical flow-based method and 2) decoding these dynamics onto music tokens with a fine-grained temporal alignment. 
DyViM captures subtle and variable dynamics cues by adapting an optical flow-based method~\cite{RAFT, VideoFlow}, without requiring any domain knowledge or specific deign catering to the type of dynamics in the video.
Music waveforms are encoded into discrete tokens with a residual vector quantizer (RVQ) model (\ie, Encodec~\cite{EnCodec}). 
To achieve temporal synchronization, DyViM interpolates the dynamics features onto these tokens, creating a continuous alignment between dynamics shifts and the music generation. 
This approach allows the music to respond in real time to the visual dynamics, providing a finer token-level synchronization.
In parallel to dynamics modeling, DyViM typically utilizes a pre-trained image encoder (\ie, CLIP~\cite{CLIP}) to extract high-level video semantics from keyframes. These features guide the music generation via cross-attention, ensuring thematic coherence. 
Additionally, an annealing tuning strategy is introduced to reduce over-constraint to the pre-trained music decoder, allowing more seamless adaptation.
Extensive experiments on three datasets validate DyViM’s effectiveness, demonstrating its ability to generate music that aligns closely with both dynamic and semantic cues, advancing the field of video-to-music generation.
Overall, our contributions are threefold:

\begin{itemize}[leftmargin=*]
    \item We underscore the crucial role of fine-grained visual dynamics in video-to-music generation, addressing an unexplored gap in leveraging nuanced cues for temporally synchronized music.
    \item We introduce a novel framework, DyViM, for video-to-music generation that enhances dynamics modeling through a specialized dynamics encoder and token-level dynamics-conditioned generation. Moreover, an annealing tuning strategy is introduced to optimize DyViM.
    \item We conduct extensive experiments across multiple datasets demonstrate DyViM’s superiority, with code and demos provided to support future research.
\end{itemize}

\begin{figure*}[t]
    \centering
    \includegraphics[trim=0 1.2cm 0 0, clip,  width=0.99\linewidth]{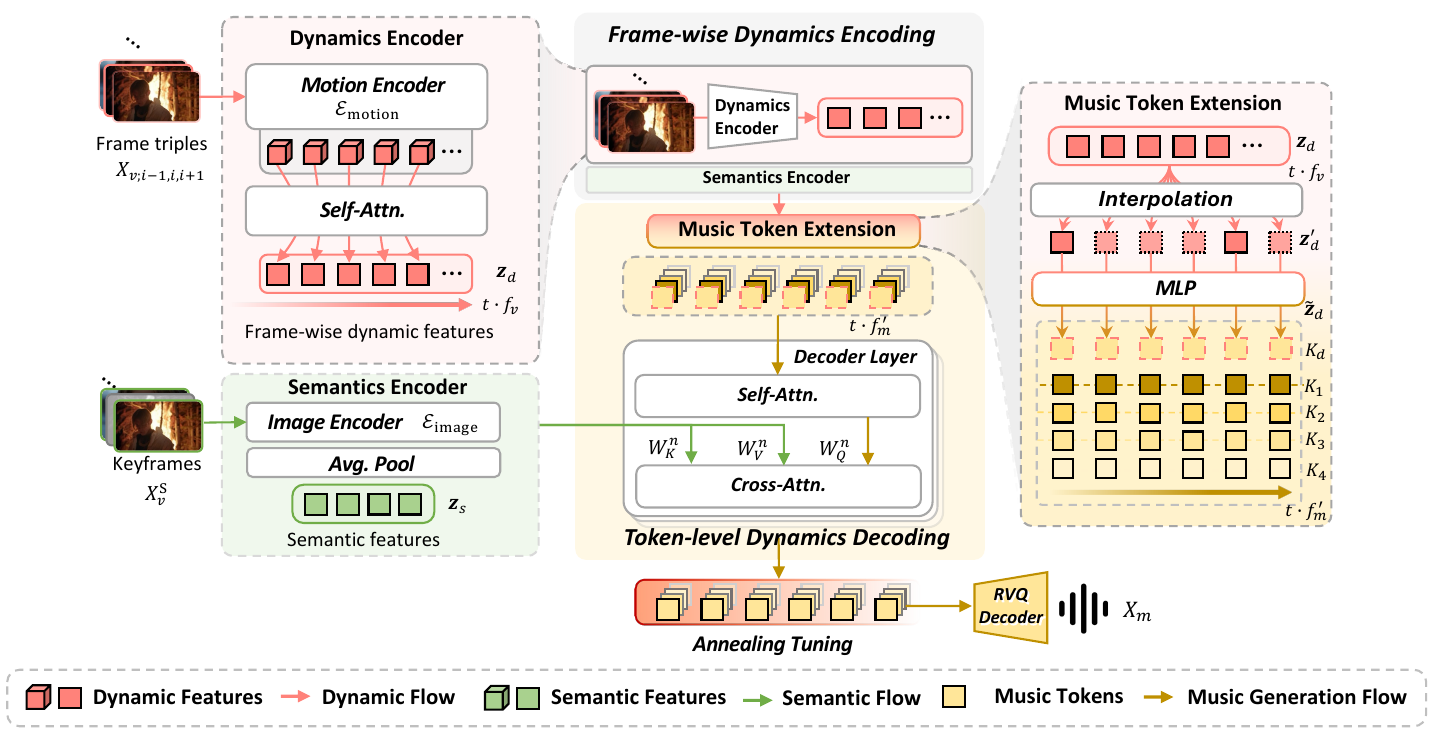}
    \caption{The overall framework of DyViM. We enhance the dynamics modeling by extracting frame-wise \cdynamic{dynamics} features and interpolating them to extend \cmusic{music} tokens to achieve token-level conditioning. 
    Additionally, keyframes provide \csemantic{semantics} with cross-attention. An annealing tuning strategy is employed to optimize the model in an effective and efficient manner. }
    \label{fig:framework}
\end{figure*}

\section{Related Work}
We review the literature on video understanding, music generation, and video-to-music generation. 

\subsection{Video Understanding}
Video understanding, a prerequisite for video-to-music generation, has advanced with improvements in representation learning and Multimodal Large Language Models (MLLMs)~\cite{Flamingo, MV-GPT, PLLaVA, VidLN}.
In video representation learning, early approaches~\cite{C3D, I3D} utilize 3D CNNs to extract spatiotemporal features 
and capture different levels of dynamics via dual-pathway architectures~\cite{SlowFast}.
Concurrently, other works utilize optical flow prediction as a self-supervised objective to model object dynamics~\cite{RAFT, VideoFlow, TransFlow}. 
More recently, transformer-based models~\cite{ViViT, VideoSwinTransformer, SwinTransformer2, SwinBert, TransFlow, su2023enhancing,10.1145/3581783.3611723} capture spatiotemporal dependencies and learn robust representations without labels. 
With the development of MLLMs, through integrating video features with language models, video language models have significant advances in video captioning~\cite{ VPCSum, Vid2Seq, Flamingo, MV-GPT} and QA~\cite{PLLaVA, VidLN, Flamingo}.

\subsection{Music Generation.}
Music generation has evolved from unconditional models to conditional models with diverse architectures and various conditions.
Different paradigms, like GANs~\cite{MuseGAN}, transformers~\cite{Music-Transformer, MusicGen}, and diffusion models~\cite{Riffusion, Diffsound, zhou2025dreamdpo, zhou2024headstudio}, are utilized to generate music without auxiliary guidance.
Recently, researchers have started to leverage conditional models regarding various inputs. 
Some models tackle continuation and inpainting tasks~\cite{MuPT, MusicGen} by given music itself.
Others condition on text or visual inputs to align music generation with user-provided textual input~\cite{MusicLM, MusicGen}, or to leverage visual-music correlations~\cite{MELFuSION, Art2Mus}.
Notably, recent works incorporate temporal dynamics into visual conditioning to study video-conditioned music generation, which is detailed in the next subsection.

\subsection{Video-to-Music Generation.}
The video-to-music generation task can be divided into distinct sub-fields depending on the video content, where each type of video is separately formulated as a specific task. 
Early studies concentrate on generating music from instrument performance videos by mapping visual cues from playing actions to corresponding musical scores~\cite{FoleyMusic, Sight2Sound, Audeo} or spectrograms~\cite{I2S}.
Subsequent research utilizes motion cues from dance videos to generate synchronized music that reflects the dynamics and emotions of the input~\cite{D2MGAN, CDCD, LORIS}.
Recently, research has increasingly shifted its focus to music videos or trailers. For example, CMT~\cite{CMT} utilizes predefined rules to extract visual cues from videos, which is then leveraged to guide the music generation process. 
VidMuse~\cite{VidMuse} introduces a long-short-term module, which only extracts the overall video embeddings without leveraging the alignment between music and video in the dataset.
Similarly, V2Meow~\cite{V2Meow} utilizes key visual semantics of sparsely sampled video frames, but potentially does not sufficiently exploit the video's temporal features.
VMAS~\cite{VMAS} emphasizes beat synchronization between video and music by utilizing a rule-based beat extraction method and weighted autoregressive loss. However, this method is limited when the video-music alignment is of low quality.
In summary, even though various methods have been proposed for video-to-music generation, they are still limited to handle nuanced dynamics and cannot achieve fine-grained temporal synchronization during generation.

\section{Preliminary}
We present the problem formulation for video-to-music generation and introduce the backbone of our approach, an autoregressive music generation model.

\subsection{Problem Formulation}
Given a video consisting of a list of image frames $\mathbf{X}_v \in \mathbb{R}^{T_v \times C \times H \times W}$, where $C$ is the number of channels and $H$ and $W$ denotes the height and width of each frame, respectively, our goal is to learn a mapping function $M$ that transforms the video into a piece of music, represented as $\mathbf{x}_m \in \mathbb{R}^{T_m}$. Here, $T_v$ and $T_m$ represents the number of frames for the video and the music, respectively.
The video and music have the same length of duration, while they have different frame rates, with $T_v = t \cdot f_v$ and $T_m = t \cdot f_m$, where $t$ is the duration, $f_v$ is the frame rate for the video, and $f_m$ is the frame rate of the music.
The mapping function is formally presented as $M: \mathbb{R}^{T_v \times C \times H \times W} \to \mathbb{R}^{T_m}$.

\subsection{Autoregressive Music Generation}
\label{sec:realted_work2}
We focus on generating the music tokens autoregressively. Here music generation models focus on using a convolutional autoencoder to generate quantized music codes~\footnote{In language modeling, codes are generally defined as tokens, serving as discrete representations.}, followed by an autoregressive decoder that generates new music tokens. 
The generation of the tokens are conditioned on either textual descriptions or a sequence of provided tokens. 
Hence, we separate the music generation process into subsequet phases: 1) Music Tokenization, and 2) Autoregressive Generation.

\subsubsection{Music Tokenization.}
In this work, we utilize a convolutional autoencoder, specifically EnCodec~\cite{EnCodec}, where the latent representation is processed through Residual Vector Quantization (RVQ)~\cite{SoundStream}. For an input music signal $\mathbf{x}_m \in \mathbb{R}^{t \cdot f_m}$, sampled at a high frequency, EnCodec transforms it into a compact latent vector with a reduced frame rate $f_m'$, where $f_m' \ll f_m$. This latent vector is subsequently quantized into a discrete set $Q \in \{1, \dots, M\}^{t \cdot f'm \times K}$, with $K$ denoting the number of codebooks and $M$ indicating the codebook size in the RVQ framework. Consequently, the $i$-th music token is derived as $\mathbf{z}_i = \sum_{k=1}^{K} \mathbf{z}_{i;k}$, aggregating contributions across all codebooks.
Leveraging the residual tokens being summed per timestep in RVQ, we extend it by incorporating visual dynamics as additional tokens. Details are provided in Section~\ref{sec:dynamics_encoder}.

\subsubsection{Autoregressive Generation.}
The autoregressive music generator models the conditional probability distribution of the next music token given the preceding tokens, formally represented as $p(\tilde{Q}_{i} \mid \tilde{Q}_{i-1}, \dots, \tilde{Q}_{0})$, where $\tilde{Q}_0$ is initialized as 0 and $i > 0$. 
To model this distribution, we adopt a music decoder $\mathcal{D}_m$, a single-stage language model, for music generation. 
To preserve musical harmony and consistency, we take advantage of the pre-tained MusicGen~\cite{MusicGen} and use it to initialize our model, instead of training the generation model from scratch.

\section{Our Approach}

We introduce DyViM to enhance dynamics modeling in video-to-music generation, as illustrated in Figure~\ref{fig:framework}. 
DyViM consists of two main modules: frame-wise dynamics feature encoding and token-level dynamics decoding. 
In addition, we integrate semantics features via a cross-attention within the decoder layers, and we employ an annealing tuning strategy to optimize the overall framework. 

\subsection{Frame-wise Dynamics Encoding}
We aim to enhance the frame-wise dynamics by designing a dynamics encoder inspired by an optical-flow method.
A self-attention module is utilized to aggregate features within frames to produce dense dynamic features.
Additionally, we follow the previous methods to extract semantics features from pre-trained image encoder~\cite{V2Meow}, to complements the dynamics features. 

\subsubsection{Dynamics Encoder.}
\label{sec:dynamics_encoder}
We encode dynamic features on a frame-wise basis to capture extensive visual motions across diverse video scenarios, including both dance and music videos. 
Inspired by previous works~\cite{RAFT, TransFlow, VideoFlow} that estimate optical flow by comparing neighboring frames, we incorporate a dynamic encoder to capture the dynamic details across frames, denoted as:
\begin{equation}
    \mathbf{z}_d^{i} = \text{DE}(\textbf{X}_{\text{v};i-1, i, i+1}),\ \text{DE} := \mathcal{E}_{\text{motion}} \circ  \text{Self-Attn},
\end{equation}
where $\textbf{X}_{\text{v};i-1, i, i+1}\in \mathbb{R}^{3\times C \times H\times W}$ represents the triplet of frames centered on the $i$-th frame. 
This setup enables us to exploit bi-directional dynamics in both forward and backward directions, following~\cite{VideoFlow}. Specifically, we elaborate on dynamics decoding with three steps. 

\noindent\textbf{Encoding motions from frame triplets.}
We calculate the correlation volumes $(\mathbf{Corr}_{i,i-1}, \mathbf{Corr}_{i,i+1})$ to measure pixel-wise visual similarity between image pairs via dot-product in downsized height $H'$ and width $W'$.
Correlation features $\mathbf{F}_{corr}^l\in\mathbb{R}^{H'\times W'\times d_{corr}}$ and flow features $\mathbf{F}_{flow}^l\in\mathbb{R}^{H'\times W'\times d_{flow}}$ are encoded from the retrieved multi-scale correlation values and predicted bi-directional flows in $l$-th refinement iteration step, respectively. 
A motion encoder is then implemented to generate motion features $\mathbf{z}_m^{i}$ through a fusion module that combines both correlation and flow information, formulated as $f : \mathbb{R}^{H\times W\times d_{c}}\times \mathbb{R}^{H\times W\times d_{f}}\to \mathbb{R}^{H\times W\times d_{m}}$. 

\noindent\textbf{Extracting music relevance with attentive aggregation.} While maintaining extensive motion details, our goal is to extract concise features that focus on music-relevant information.  
Therefore, we aggregate the generated motion features via a self-attention mechanism within frames, followed by average pooling to obtain the attentive dynamics $\mathbf{z}_d^{i}\in\mathbb{R}^{d_m}$ for each frame. 

\noindent\textbf{Organizing.}  Ultimately, the dynamic features are represented as $\mathbf{z}_d\in\mathbb{R}^{t\cdot f_v\times d_m} := [ \mathbf{z}_d^{0}, \dots, \mathbf{z}_d^{t\cdot f_v}]$.

\subsubsection{Semantics Encoder.} 
Additionally, we introduce the semantics, which can be attributed to the visual content extracted from keyframes.
A straightforward approach to capture the semantics from video is to utilize a pre-trained video captioning model, which summarizes the video content into textual descriptions~\cite{}.  
However, such methods often ignore the visual nuances when translating visual features into text.  
To extract the semantic features from video, we leverage a pre-trained 2D visual encoder $\text{SE}(\cdot)$, such as CLIP~\cite{CLIP}, to encode visual content from keyframes.  
Specifically, we first employ the MPEG-4~\cite{MPEG-4} compression technique to extract keyframes, where I-frames in MPEG-4 are used~\cite{Video-LaVIT}. These frames are denoted as $\mathbf{X}_{\text{v}}^{\text{S}}\in\mathbb{R}^{N_s\times C \times H\times W}$, where $N_s$ represents the number of keyframes.  
Subsequently, we employ a semantic encoder to obtain semantic features, denoted as:
\begin{equation}
     \mathbf{z}_s = \text{SE}(\mathbf{X}_{\text{v}}^{\text{S}}),\ \text{SE} := \mathcal{E}_{\text{image}} \circ \text{Avg-Pool}, 
\end{equation}
where $\mathcal{E}_{\text{image}} :\mathbb{R}^{N_s\times C \times H\times W} \to \mathbb{R}^{N_s\times N_h \times d_s}$ transforms the $N_s$ images into $d_s$-dimensional latent features, with $N_h$ being the number of hidden states.  
We then apply a simple average pooling operation, $\text{Avg-Pool}: \mathbb{R}^{N_s\times N_h \times d_s} \to \mathbb{R}^{N_h \times d_s}$, to efficiently aggregate these features.

\subsection{Token-level Dynamics Decoding}
To effectively generate music from nuanced visual features, we interpolate dynamics features to extend music tokens, achieving fine-grained temporal synchronized conditioning. 
And semantics are integrated via cross-attention modules. 

\subsubsection{Music Token Extension.}
Visual dynamics offer detailed conditions for music generation, introducing nuanced variations that facilitate rhythmic and temporal synchronization, providing low-level control. 
However, recent methods either apply visual dynamics as a global control~\cite{D2MGAN, LORIS} or utilize direct matching (\eg, onsets~\cite{VMAS} or optical flow magnitude~\cite{Video2Music}) for strict synchronization, which limits fine-grained and flexible conditioning.
Inspired by the intrinsic structure of music codes, where four codes represent a single music token, where each code complements the prior ones, as outlined in Section~\ref{sec:realted_work2}—we propose integrating dynamic shifting into these existing tokens with the following two alignments.

\noindent\textbf{Frame-level alignment.}
We first interpolate the dynamics features to match the number of music frames, thus yielding $\mathbf{z}'_d \in \mathbb{R}^{t\cdot f_m\times d_m}$. 
We adopt a fast and straightforward way of nearest-neighbor interpolation to find the closest dynamics for new points, where $\mathbf{\tilde{z}}'_{d;i'} = \mathbf{\tilde{z}}_{d;i}$, $i' = \text{round}(t\cdot\frac{f_m}{f_v}\cdot i)$.

\noindent\textbf{Dimensional alignment.} We transform these features to align with the dimensionality of the music features using a Multilayer Perceptron (MLP), producing $\mathbf{\tilde{z}}_d\in \mathbb{R}^{t\cdot f_m\times d} = \text{MLP}(\mathbf{z}'_d)$.  
To incorporate the transformed dynamic features, we extend the existing music codes derived from RVQ codebooks with interpolated dynamic features and sum them up to obtain the music tokens, defined as:
\begin{equation}
    \mathbf{z}_i := \text{sum}(\alpha\cdot \mathbf{\tilde{z}}_{d;i}, \{(1-\alpha)\cdot\mathbf{z}_{i;k}; k \in [K]\}),
    \label{eq:dynamic}
\end{equation}
where $\mathbf{\tilde{z}}_{d;i}$ represents the dynamic feature, and $\{\mathbf{z}_{i;k}\}$ represents the set of music codes from the codebooks, with $k$ indexing over the set $[K]$; and $\alpha$ controls the strength of dynamics to the music codes. 

Ultimately, music codes extension enables fine-grained control over music generation by embedding dynamic visual features, allowing the generated music to reflect subtle visual changes accurately.

\subsubsection{Semantics Condition.}
Visual semantics provide fundamental guidance for music generation, maintained throughout the entire process. 
To this end, we propose incorporating visual semantics as high-level signals, which are temporally invariant and control the foundational tones of the music.  
Similarly, we adopt an MLP projector, to transform visual semantics into music conditions, denoted as $\mathbf{\tilde{z}}_s\in \mathbb{R}^{N_d\times d} = \text{MLP}(\mathbf{z}_s)$, thus maintaining dimensional consistency~\cite{Llava, MiniGPT-4}. 
To condition the music generation, we employ a cross-attention mechanism, denoted as:  
\begin{equation}
    \begin{aligned}
    \mathbf{A}^{n} &= \frac{1}{\sqrt{d}}\mathbf{z}^{n-1}\mathbf{W_q^n}(\mathbf{\tilde{z}}_s\mathbf{W_k^n})^\top, \\
    \mathbf{z}^{n} &= \text{softmax}(\mathbf{A}^{n})\mathbf{\tilde{z}}_s \mathbf{W_v^n},
\end{aligned}
\end{equation}
where $\mathbf{W_q^n}$, $\mathbf{W_k^n}$, and $\mathbf{W_v^n} \in\mathbb{R}^{d \times d}$ are attention weight matrices that transform the music features $\mathbf{z}^{n-1}$ and semantic conditions into query, key, and value spaces at the $n$-th layer.  
$\mathbf{A}^{n}$ is the corresponding attention matrix, followed by a softmax function to normalize attention scores, resulting in new representations at the next layer.

\subsection{Annealing Tuning}
To leverage the effectiveness of pre-trained music generation models while preserving their inherent musical knowledge, we employ an annealing fine-tuning strategy.  
Specifically, we fine-tune the music decoder with only a small set of parameters by adopting LoRA~\cite{LoRA} with an annealing schedule.  
Following an autoregressive approach, we define the annealing tuning formally as:
\begin{equation}
    \max_{\Theta} \sum_{\tau=1}^{|Q|} 
    \log 
    \mathbf{a}_{\tau} p(\tilde{Q}_{\tau} | Q_{< \tau}; \mathbf{z}_s, \mathbf{z}_d),
    \label{eq:annealing}
\end{equation}
where $\Theta$ represents all trainable parameters, including the LoRA weights and modules from the decomposition and composition processes.  
The annealing schedule $\mathbf{a}_{\tau} \in \mathbb{R}^{t\cdot f_m}$ controls the weight of each token, with more emphasis placed on the initial tokens, and gradually reducing the weight over time.  
The rationale is that the musical theme or structure is primarily established during the critical early stages of music generation. Reducing the weight on later tokens allows the model to rely on its well-trained understanding of musical structure, enabling coherent music generation without being overly constrained by the given samples.
We employ a cosine decay schedule, defined as 
$\mathbf{a}_\tau = {a_{\text{max}} \cdot (1 + \cos(\frac{\pi \cdot \tau}{|Q|}))}/{2} + \epsilon$,
where $a_{\text{max}}$ and $\epsilon$ denote the maximum and minimum weight values, respectively. Moreover, we analyze various schedules, like linear or step decays in Section~\ref{sec:annealing_schdules} for a comprehensive evaluation. 
\section{Experimental Setup}
We elucidate the experimental setup, covering implementation details, datasets, metrics, and selected baselines to ensure a fair and comprehensive evaluation.
\begin{table*}[t]
\centering
\caption{The overall performance comparison between our DyViM and various baselines on three datasets regarding both subjective (Obj. Eval.) and objective (Subj. Eval.) evaluation metrics. Bold and underlined indicate the best and the second-best performance.}
\label{tab:overall_performance}
\resizebox{\textwidth}{!}{
\begin{tabular}{l|ccc|cc|ccc|cc|ccc|cc} 
\toprule
\textbf{Dataset} & \multicolumn{5}{c|}{\textbf{AIST++}}                   & \multicolumn{5}{c|}{\textbf{SymMV}}                  & \multicolumn{5}{c}{\textbf{BGM909}}                      \\ 
\midrule
\multirow{2}{*}{\textbf{Metrics}} & \multicolumn{3}{c|}{Obj. Eval.} &\multicolumn{2}{c|}{Subj. Eval.}    & \multicolumn{3}{c|}{Obj. Eval.} &\multicolumn{2}{c|}{Subj. Eval.}                 & \multicolumn{3}{c|}{Obj. Eval.} &\multicolumn{2}{c}{Subj. Eval.}                     \\ 
\textbf{}   & FD$\downarrow$ & {FAD$\downarrow$} & {KL$\downarrow$} & {OVL$\uparrow$} & {REL$\uparrow$} & {FD$\downarrow$} & {FAD$\downarrow$} & {KL$\downarrow$} & {OVL$\uparrow$} & {REL$\uparrow$} & {FD$\downarrow$} & {FAD$\downarrow$} & {KL$\downarrow$} & {OVL$\uparrow$} & {REL$\uparrow$}  \\ 
\midrule
\textbf{CMT}~\cite{CMT} &   63.65    &   15.59    &    1.24     & 62.05  & 49.23 &  54.86   &   7.13    &   \underline{0.87}    & 53.91   &  42.61          &  51.09     &   8.56   &  1.02      &  45.21  &   30.91        \\
\textbf{D2MGAN}~\tablefootnote[8]{D2MGAN and CDCD specialize in dance videos with only processed features provided, so we report results on their self-split AIST++ dataset.}~\cite{D2MGAN}& 48.37 & 11.40 & 0.94 & 46.67 & 58.33 &- &- &- &- &- &- &- &- &- &- \\
\textbf{CDCD}~\footnotemark[8]~\cite{CDCD} & 54.80 & 7.20 & \textbf{0.40} & 50.62 &\underline{76.25} &- &- &- &- &- &- &- &- &- &- \\
\textbf{Video2Music}~\cite{Video2Music}&  90.15 & 31.53 &  1.54 &64.10 & 43.59 &  81.06   & 20.70 &   1.03 & 60.87 & \underline{60.21} &  86.25 &  24.84  &   1.29 & {61.73}  &   51.30      \\
\textbf{DiffBGM}~\cite{Diff-BGM} &88.12 & 23.72 & 1.53 &\underline{66.40} & 37.6 &71.44 &16.23 & 0.93 &\underline{64.37} & 47.50 & 75.63 & 9.90 & 1.22 & \underline{61.87} & 44.37\\
\textbf{M$^2$UGen}~\cite{MU2Gen} & 62.89 & 18.39 &   {0.98} & 26.67 &   36.41       &  \underline{47.91} & 5.49 &  1.07 & 56.52 & 43.64 & 45.70 & 5.75 & 1.23 & 49.56  &  44.34         \\ 
\textbf{VidMuse}~\cite{VidMuse}&  \underline{46.50} &   \underline{6.03} &   0.99& 58.46 & {53.84} &  50.19       &   \underline{4.78}           &  1.03 & 60.21 & 50.43 & \underline{32.41}      &  \textbf{2.80} &  \textbf{0.93} & 60.86 & \underline{61.73}\\
\midrule
\rowcolor[rgb]{0.89,0.922,0.988} \textbf{DyViM (ours)}   & \textbf{26.86}  &   \textbf{4.11}          &  \underline{0.75}         & \textbf{74.35}  &         \textbf{77.43} &   \textbf{31.99}       &   \textbf{3.24}           &    \textbf{0.77}       & \textbf{71.30} &  \textbf{76.52}       &  \textbf{24.19}       &   \underline{3.40}           &   \underline{0.94}        & \textbf{62.61} &    \textbf{63.48}     \\ 
\bottomrule
\end{tabular}
}
\end{table*}

\subsection{Implementation Details}

\subsubsection{Visual Encoder.} 
We adapt the official VideoFlow library~\footnote{\url{https://github.com/XiaoyuShi97/VideoFlow}} and extract the motion features ($d_m$ = 1024) for dynamic feature aggregation, and employ the official CLIP library~\footnote{\url{https://github.com/openai/CLIP}} and extract the last hidden states ($N_h$ = 50 and $d_s$ = 768) for semantic feature aggregation. 

\subsubsection{Music Decoder.} We adopt MusicGen~\cite{MusicGen} as the music decoder, and EnCodec~\cite{EnCodec} as the compression model that tokenize waveforms, where the dimension of music features is 1536, \ie, $d$=1536. Their official library~\footnote{\url{https://github.com/facebookresearch/audiocraft}} is adapted for further modification. 
MusicGen is equipped with 48 transformer layers, followed by 4 linear layer to map music features to the indexes of music codes. 
While EnCodec is a convolutional neural network, compresses  1-second waveform to 50 discrete audio tokens within four codebooks with the size of 2048.

\subsubsection{Hyper-parameter Setting.} 
We split the video in to 10-second clips to form the datasets. 
And we utilize the AdamW optimizer ($\beta_1$=0.9 and $\beta_2$=0.95) with a batch size of 8 and warming up for 100 steps in a cosine learning schedule. 
Notably, our training is efficient with a single 48GB A40 GPU device.

\subsection{Datasets}
We follow the prior video-to-music methods~\cite{CMT,LORIS,Diff-BGM,D2MGAN}, while evaluating on more diverse datasets to evaluate the efficacy of proposed methods. 
Specifically, the collected dataset incorporates two categories: dance and music videos. 
The dance videos are curated from AIST++~\cite{AIST++}, with clear visual pacing accompanies rhythmic songs.
The music videos refers to the a short film, mostly focusing on the the content to visual storytelling.
We collect this type of videos from SymMv~\cite{V-MusProd} and BGM909~\footnote{The BGM909 dataset changes the original music tracks of the videos using midi music retrieved from POP909. In our case, we choose to keep the original version to maintain fidelity.}~\cite{Diff-BGM}. 
And we split the dataset for training and testing with the ratio of 9:1 to ensure fairness and avoids the information leakage. 

\subsection{Evaluation Metrics}
For a comprehensive evaluation, we utilize both objective and subjective metrics to evaluate the overall performance. 

\subsubsection{Objective Evaluation.} contains the estimations between generated music to the real music in associated video.  
Following previous work~\cite{V2Meow, MU2Gen, VidMuse}, we compute the Fréchet Distance (FD), Fréchet Audio Distance (FAD)~\cite{FAD} and Kullback-Leibler Divergence (KL)~\cite{KL}.
Wherein, FAD compares the statistical distribution between generated and real music via their extracted high-level feature from VGGish~\cite{VGGish}, while FD uses PANNs~\cite{PANNs} as the feature extractor.
And KL compute the divergence of labels between generated and original music. 
These evaluation can be implemented via well-established libraries~\footnote{\url{https://github.com/haoheliu/audioldm\_eval}}.

\subsubsection{Subjective Evaluation.} Inspired by recent text/image-to-music methods~\cite{MELFuSION, AudioGen}, we assess the generated samples using two metrics: overall quality (OVL) and relevance to the input video (REL), each rated on a range from 1 to 10~\footnote{The reported scores are scaled up to 100 for intuitive presentation.}.

\subsection{Baselines}
To make an exhaustive evaluation, we choose several well performed and accessible methods as baselines. 
\textbf{CMT}~\cite{CMT} pre-defines visual features to connect video to music via motion and timing and generate symbolic music. 
\textbf{D2MGAN}~\cite{D2MGAN} is specialized for dance videos with a GAN-based discriminator for generating waveforms with human body motions.
\textbf{CDCD}~\cite{CDCD} advances D2MGAN by combining diffusion and constrative learning for dance-to-music generation.
\textbf{Video2Music}~\cite{Video2Music} leverages semantic, motion and emotional features to predict symbolic music by training a specialized transformer decoder. 
\textbf{DiffBGM}~\cite{Diff-BGM} segments input video for frames and captions, followed with visual and language encoders and output piano roll conditioned with a diffusion model. 
\textbf{M$^2$UGen}~\cite{MU2Gen} employs a LLM to understand multiple modalities (\eg, video and text) and generates waveform music via a frozen pre-trained music decoder. 
\textbf{VidMuse}~\cite{VidMuse} utilizes a long-short-term visual module to obtain visual features and trains a pre-trained music decoder with extensive data.
Notably, we implement these baselines rigidly following their officially provided code to ensure their promised performance. 

\section{Results and Analysis}

To demonstrate the effectiveness and rationale of DyViM, we conduct extensive experiments, including an overall comparison with baselines, ablation study on conditioning and tuning strategies, and model analyses on dynamics control, visual feature selection, and annealing schedules.

\subsection{Overall Performance}
To demonstrate the effectiveness of DyViM, we present both objective and subjective comparisons across three datasets, as shown in Table~\ref{tab:overall_performance}. 
DyViM consistently outperforms the baselines, indicating its superiority in generating both high-fidelity and visually synchronized music. 
Specifically, we have multiple observations.
First, DyViM significantly improves upon the baselines, even surpassing specialized models on certain metrics, such as D2MGAN and CDCD for dance-to-music generation, and VidMuse, which was trained on 200K music videos.
Second, symbolic music generation yields lower objective scores due to substantial distributional gaps with test datasets, but provides a relatively comfortable listening experience in subjective evaluations due to its generation policy based on symbolics, as exemplified by Video2Music and Diff-BGM.
Third, both M$^2$UGen and VidMuse leverage pre-trained music decoders (\eg, MusicGen), leading to improvements in objective scores. 
Finally, the paradigm of freezing the music decoder (\ie, M$^2$UGen) causes disharmony, significantly disrupting the listening experience and leading to lower subjective scores.
Both objective and human evaluations compared with baselines demonstrate the effectiveness of DyViM.

\subsection{Ablation Study}

\subsubsection{Impact of different conditions.}
To verify the importance of different conditions (\ie, dynamics, denoted as \textbf{D}, and semantics, denoted as \textbf{S}), we perform ablations to create different model variants, as shown in Table~\ref{tab:visual_conditions}. 
The results showcases a significant performance drop when dynamics are excluded, compared to the case without semantics. This might indicate that visual dynamics provide more informative guidance than high-level semantics in the music generation process.

\subsubsection{Impact of different condition methods.}
To explore various methods for conditioning music generation from video, we also replace the conditioning methods in the two-level composition process as shown in Table~\ref{tab:visual_conditions}.
Specifically, we test: 1) \text{cross}, which uses a cross-attention mechanism; 2) \text{prepend}, which adds the features as prefix tokens; and 3) \text{extend}, representing our proposed method of extending music codes.
Among these conditioning methods, our proposed approach of extending music codes combined with semantic conditioning via cross-attention achieves the best performance. This result further supports the rationale behind our design, demonstrating its effectiveness in adapting to different levels of visual cues.

\begin{table}
\centering
\caption{Performance comparison \textit{w.r.t.} different visual conditions and strategies.}
\label{tab:visual_conditions}
\resizebox{\linewidth}{!}{
\begin{tabular}{l|cc|cc|cc} 
\toprule
                 & \multicolumn{2}{c|}{Visual conditions} & \multicolumn{4}{c}{Condition strategies}       \\
\textbf{Dataset} & \textbf{w/o-D} &\textbf{w/o-S}                          & \textbf{D}$_\text{cross}$& \textbf{D}$^*_\text{extend}$ & \textbf{S}$_\text{prepend}$ & \textbf{S}$^*_\text{cross}$   \\ 
\midrule
\textbf{AIST++}  & 4.63  & 4.59                           & 4.41     & 4.11      & 4.45       & 4.11        \\
\textbf{SymMV}   & 4.02  & 3.67                           & 3.57     & 3.24      & 4.16       & 3.24        \\
\textbf{BGM909}  & 4.39  & 4.29                           & 4.26     & 3.40      & 4.49       & 3.40        \\
\bottomrule
\end{tabular}
}
\end{table}

\subsubsection{Performance regarding to different tuning strategies.}
\begin{figure}
    \centering
    \includegraphics[width=\linewidth]{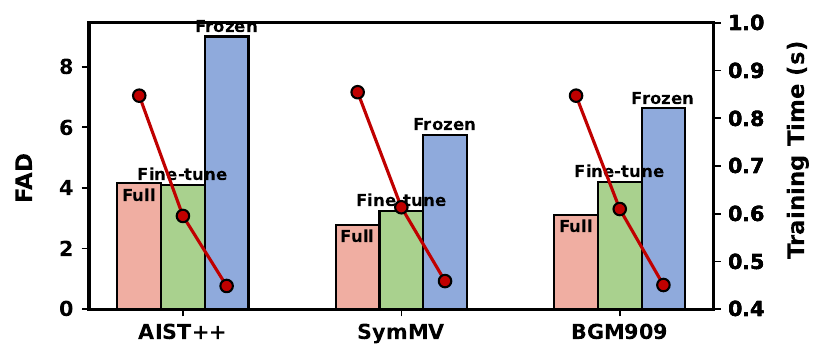}
    \caption{Performance comparison (bar) and the time cost during training (line) \textit{w.r.t.} different tuning strategies.}
    \label{fig:tuning}
\end{figure}

We conduct performance and training time comparisons across different tuning strategies, as shown in Figure~\ref{fig:tuning}, to demonstrate the effectiveness and efficiency of our fine-tuning strategy.
Wherein, \textbf{Full} denotes updating all parameters, \textbf{Fine-tune} represents our approach that trains only a subset of parameters, and \textbf{Frozen} keeps all parameters fixed.
The results reveal a clear performance hierarchy: $\textbf{Full} \geq \textbf{Fine-tune} > \textbf{Frozen}$, and an efficiency hierarchy: $\textbf{Full} < \textbf{Fine-tune} < \textbf{Frozen}$. The fine-tuning approach achieves an effective trade-off between performance and efficiency. Notably, the full tuning setting incorporates our annealing schedule, which helps mitigate overfitting, leading to enhanced performance.

\subsection{Model Study}
\begin{figure}
    \centering
    \includegraphics[width=\linewidth]{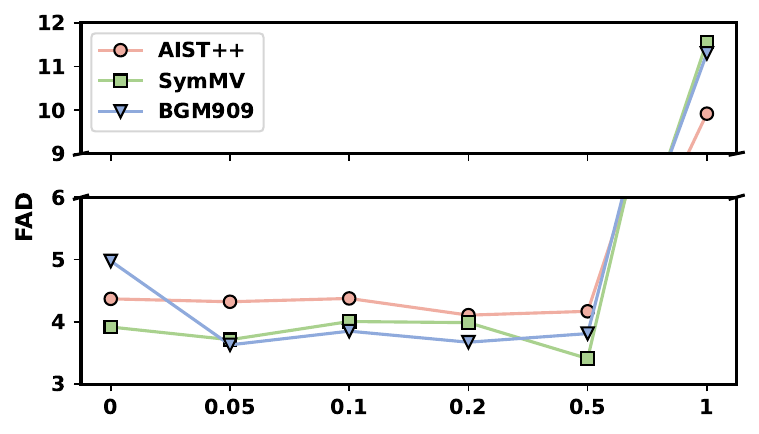}
    \caption{Performance comparison \textit{w.r.t.} different $\alpha$ to control the strength of visual dynamics.}
    \label{fig:alpha}
\end{figure}

\subsubsection{Impact of $\alpha$ to control dynamics.}
The parameter $\alpha$, introduced in Equation~\ref{eq:dynamic}, represents the strength of the dynamic feature when extending the music tokens. 
To illustrate its impact, we vary its value within the range [0, 1] and present a performance comparison in Figure~\ref{fig:alpha}. 
Incorporating visual dynamics generally results in lower FAD scores (\ie, improved performance), while the extreme cases of $\alpha=0$ and $\alpha=1$ yield poorer results. 
In our experiments, a larger value of $\alpha$ tends to destabilize the learning process and is prone to convergence to a suboptimal state. 
Therefore, we emphasize tuning $\alpha$ below 0.5. 

\subsubsection{Impact of different visual encoders.}
We replace different visual encoders for both dynamic and semantic encoding to investigate their impacts, with performance comparisons shown in Table~\ref{tab:visual_encoders}. 
For dynamics, we compare our proposed method (\ie, \textbf{D}$^\text{FwDF}$) with OpenPose~\cite{OpenPose}, which is specifically used for dance videos~\cite{D2MGAN, CDCD, LORIS}, and I3D~\cite{I3D}, which also leverages optical flow information but with frame compression. Compared to other dynamic encoders, our method demonstrates superior performance, which can be attributed to its capability to capture nuanced visual information and attentively guide music generation.
For semantics, we employ a video captioning model (\ie, PLLaVA~\cite{PLLaVA}) to obtain descriptive captions, followed by a T5 model~\cite{T5}.  
The transformation from video $\to$ language $\to$ music may introduce information loss, which limits its performance compared to a direct image encoder like CLIP~\cite{CLIP}, especially on datasets like SymMV and BGM909, which contain complex semantics. 
However, for simpler semantics, such as in AIST++, which consists mainly of dance scenes, T5-based approach slightly outperforms visual encoder methods.

\begin{table}
\centering
\caption{Performance comparison \textit{w.r.t.} different visual encoders.}
\label{tab:visual_encoders}
\resizebox{\linewidth}{!}{
\begin{tabular}{l|ccc|cc} 
\toprule
                 & \multicolumn{3}{c|}{Dynamic encoder} & \multicolumn{2}{c}{Semantic encoder}  \\
\textbf{Dataset} & DE$^\text{OpenPose}\tablefootnote[9]{OpenPose specializes in extracting human body motions, making it suitable only for the AIST++ dataset.}$ & DE$^\text{I3D}$& DE$^\text{FwDF*}$ & SE$^\text{T5}$ & SE$^\text{CLIP*}$                   \\ 
\midrule
\textbf{AIST++}  &   4.48     & 4.23     & 4.11      & 4.09    & 4.11                        \\
\textbf{SymMV}   & -             & 5.31     & 3.24      & 4.29    & 3.24                        \\
\textbf{BGM909}  & -             & 3.85     & 3.40      & 4.28    & 3.40                        \\
\bottomrule
\end{tabular}
}
\end{table}

\subsubsection{Dynamics attention.}
\begin{figure}[h]
    \centering
    \includegraphics[width=\linewidth]{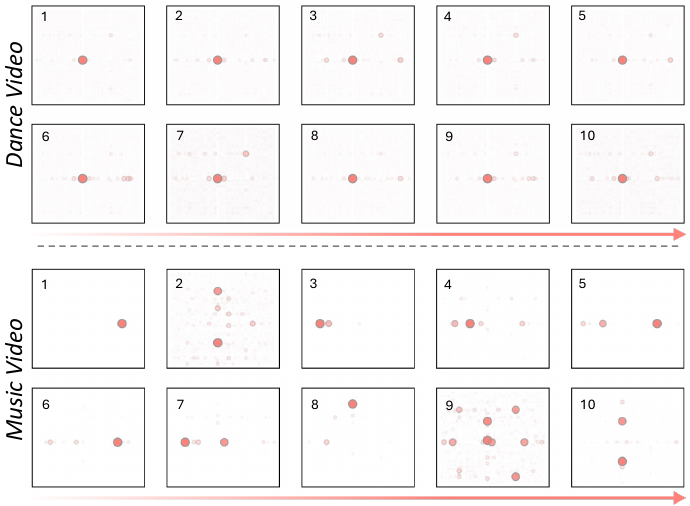}
    \caption{An illustration of the dynamics attention matrix for dance videos (top) and music videos (bottom), where the size of each point represents the quantitative value of attention.}
    \label{fig:attn_matrix}
\end{figure}

We illustrate the attention matrix generated by the self-attention modules in the dynamics encoder, as shown in Figure~\ref{fig:attn_matrix}.  
There are distinct attention patterns for different types of videos. For intuitive visualization, we select 10 frames.  
For dance videos, the dynamics attention is relatively focused, which is reasonable since dance videos typically feature a fixed object within the scene producing motions.  
In contrast, music videos, which exhibit more diverse dynamics, display varying attention patterns across frames.  
Our proposed simple self-attention module, trained with token-level dynamics decoding, effectively captures these nuanced dynamic changes, thereby enhancing finer video-to-music generation.

\subsubsection{Case studies on visual dynamics.}
\begin{figure}
    \centering
    \includegraphics[width=\linewidth]{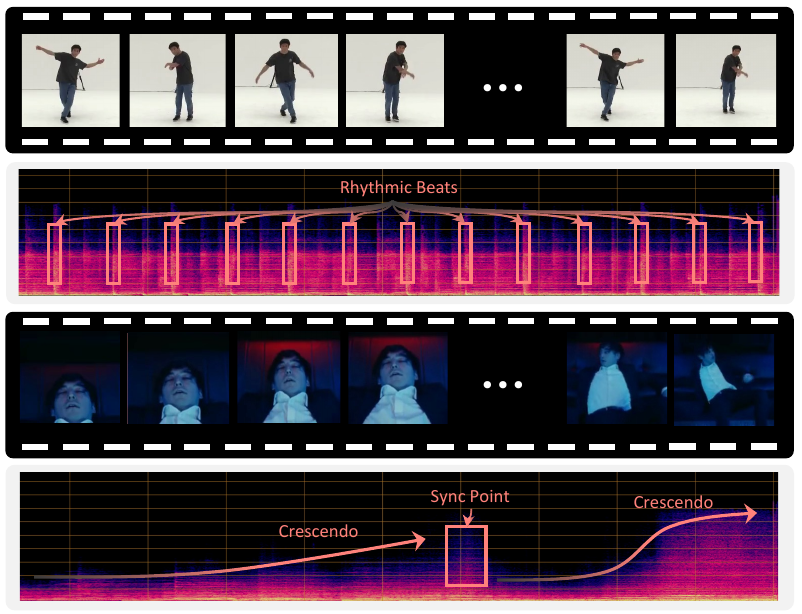}
    \caption{Case study on generated music samples from dance and music videos.}
    \label{fig:case_study}
\end{figure}

We analyze the dynamic visual correlations between the generated music and associated videos, as illustrated in Figure~\ref{fig:case_study}. Notably, DyViM produces visually relevant music that aligns with visual dynamics (\eg, body motions corresponding to rhythmic beats and camera movements to music crescendos). More cases are available in the Demo and supplementary materials.

\subsubsection{Different annealing schedules.}
\label{sec:annealing_schdules}
Annealing tuning is central to adapting the autoregressive music decoder for video conditioning in DyViM. The main idea is to prioritize the initial tokens while gradually reducing the weights of later tokens. We examine five different annealing schedules in Table~\ref{tab:annealing_schedules}: 
1) $\mathbf{a}^\text{constant}_\tau = 1$;  
2) $\mathbf{a}^\text{random}_\tau \sim \text{Uniform}(\epsilon, a_{\text{max}})$;  
3) $\mathbf{a}^\text{step}_\tau = \mathbb{I}(\tau<|Q|/2) \cdot a_{\text{max}} + \epsilon$;  
4) $\mathbf{a}^\text{linear}_\tau = \frac{(|Q| - \tau) \cdot a_{\text{max}}}{|Q|} + \epsilon$;  
5) $\mathbf{a}^\text{cosine}_\tau = \frac{a_{\text{max}} \cdot (1 + \cos(\frac{\pi \cdot \tau}{|Q|}))}{2} + \epsilon$. 
We consider $\mathbf{a}^\text{constant}$ and $\mathbf{a}^\text{random}$ as non-annealing tuning variants, with $\mathbf{a}^\text{constant}$ representing the most conventional fine-tuning method. 
The results clearly showcase that incorporating an annealing schedule significantly improves performance, further validating the rationale and effectiveness of our approach.

\begin{table}
\centering
\caption{Performance comparison \textit{w.r.t.} different tuning schedules.}

\label{tab:annealing_schedules}
\resizebox{\linewidth}{!}{
\begin{tabular}{l|cc|ccc} 
\toprule
                 & \multicolumn{2}{c|}{w/o-annealing} & \multicolumn{3}{c}{w/-annealing}       \\
\textbf{Dataset} & $\mathbf{a}^\text{constant}$ & $\mathbf{a}^\text{random}$      & $\mathbf{a}^{\text{step}}$ & $\mathbf{a}^{\text{linear}}$ &$\mathbf{a}^{\text{cosine*}}$ \\ 
\midrule
\textbf{AIST++}  & 4.87          & 4.76               & 4.18      & 4.24        & 4.11         \\
\textbf{SymMV}   & 4.11          & 4.43               & 3.39      & 3.32        & 3.24         \\
\textbf{BGM909}  & 4.64          & 4.71               & 3.59      & 3.64        & 3.40         \\
\bottomrule
\end{tabular}
}
\end{table}

\section{Conclusion}
In this paper, we emphasized the significance of visual dynamics for video-to-music, while exhibiting an unexplored gap in leveraging nuanced dynamic cues for temporally synchronized music. 
To this end, we presented DyViM, a novel framework that enhances dynamics modeling in video-to-music generation.
Specifically, we encoded the frame-wise dynamics features by adapting optical-flow based methods to capture motion features across frame triples, followed with a self-attention to aggregate music-relevant dynamics within frames.
These features are then decoded onto music tokens, adding dynamics shifts to achieve fine-grained temporal synchronization. 
Moreover, semantics are incorporated with cross-attention for high-level conditions. 
We proposed an annealing tuning strategies to facilitate the model optimization.
We conducted extensive experiments, comparing DyViM with SOTA methods across three datasets and evaluating results through both objective and subjective metrics. 
We hope to push the boundaries of this field, especially by delving into the complex dynamics between video and music and providing capabilities that work with it out of the box.
In future work, we will explore multiple meaningful directions, including:
1)
Introducing comprehensive evaluation dimensions, including new metrics, novel benchmarks, etc., for holistic video-to-music assessment;
2)
Handling long-sequence video-to-music generation that maintains temporal consistency while incorporating narrative progression and structural variation over extended durations;
3)
Investigating interactive video-to-music generation systems that allow diverse forms of user instructions and customization.

\bibliographystyle{ACM-Reference-Format}
\balance
\bibliography{biblio}

\appendix
\clearpage
\setcounter{page}{1}

\section{Data Pre-Processing Details}

We summarize the statistics of the datasets in Table~\ref{tab:data_statistics}. The datasets were downloaded from their official repositories: \texttt{AIST++}\footnote{\url{https://github.com/L-YeZhu/D2M-GAN}}, \texttt{SymMV}\footnote{\url{https://github.com/zhuole1025/SymMV}}, and \texttt{Diff-BGM}\footnote{\url{https://github.com/sizhelee/Diff-BGM}}. To preserve the original video-music correlations, we retain the waveform tracks provided in the datasets rather than adopting symbolic music (\eg, MIDI) or substituted tracks (\eg, POP909 in Diff-BGM).
\begin{table}[h]
\centering
\caption{Statistics of datasets.}
\label{tab:data_statistics}
\resizebox{\linewidth}{!}{
\begin{tabular}{lccc} 
\toprule
       & Video Content & Size  & Length (Hours)  \\ 
\midrule
AIST++ & Dance Video   & 1,408 & 5.2             \\
SymMV  & Music Video   & 1,140 & 76.5            \\
BGM909 & Music Video   & 909   & 62.6     \\
\bottomrule
\end{tabular}
}
\end{table}

For training consistency and generation quality, we apply the following pre-processing steps: 1) Vocal Removal: Using \texttt{vocal-remover}~\footnote{\url{https://github.com/tsurumeso/vocal-remover}}, we remove vocals to focus on instrumental tracks. 2) Silence Detection: Prolonged silences are detected and removed with \texttt{pydub}\footnote{\url{https://github.com/jiaaro/pydub}} to ensure consistent audio energy.
These steps maintain the integrity of the datasets while facilitating training and generation quality.

\section{Implementation Details of Baselines}

To ensure a rigorous and fair comparison, we carefully implement and adapt the baseline models following the instructions and publicly available codebases. 
\textbf{CMT}~\cite{CMT} and \textbf{Video2Music}~\cite{Video2Music} are models that generate symbolic music based on pre-defined visual features and rule-based broken chord approach, respectively, both using official implementations and pre-trained weights\footnote{\url{https://github.com/CMT-repository}}\footnote{\url{https://github.com/Video2Music-repository}}. \textbf{DiffBGM}~\cite{Diff-BGM}\footnotemark[3] uses diffusion-based generation with visual and language encoders, relied on multiple open-source feature extractors: VideoCLIP\footnote{\url{https://github.com/CryhanFang/CLIP2Video}} for video features, BLIP\footnote{\url{https://github.com/salesforce/BLIP}} for video captions, Bert-base-uncased\footnote{\url{https://huggingface.co/google-bert/bert-base-uncased}} as the language encoder, and TransNetV2\footnote{\url{https://github.com/soCzech/TransNetV2}} for shot detection. We integrated these extractors with their default setting. 
Since the three aforementioned models produce symbolic music as outputs, we synthesized the results into audio using FluidSynth\footnote{\url{https://www.fluidsynth.org/}} and the FluidR3\_GM soundfont\footnote{\url{https://musical-artifacts.com/artifacts/738}}.
Notably, DiffBGM's audio output was trimmed to match the video duration by retaining the initial segment.
\textbf{D2MGAN}~\cite{D2MGAN} and \textbf{CDCD}~\cite{CDCD} are Dance2Music approaches heavily relying on OpenPose features, limiting their application to datasets like SymMV and BGM909. Both of them are trained on AIST++ and provide only pre-segmented clips and extracted features without preprocessing scripts; we utilized their provided model weights and calculated scores only on their test sets to avoid inconsistencies in dataset splits, using the 2-second segmentation for D2MGAN and 6-second segmentation for CDCD, as recommended in their repositories\footnote{\url{https://github.com/D2MGAN-repository}, \url{ https://github.com/L-YeZhu/CDCD}}. 
\textbf{M$^2$UGen}~\cite{MU2Gen}\footnote{\url{https://github.com/M2UGen-repository}} and \textbf{VidMuse}~\cite{VidMuse}\footnote{\url{https://github.com/VidMuse-repository}} are end-to-end models capable of generating raw music directly from video inputs. 
Although their official repositories did not provide batch inference scripts, they included Gradio-based\footnote{\url{https://www.gradio.app/docs}} demo interfaces. We modified these demos to develop custom inference scripts to evaluate these models on our test datasets. 

For \textbf{VMAS}~\cite{VMAS}\footnote{\url{https://genjib.github.io/project_page/VMAS/index.html}}, \textbf{LORIS}~\cite{LORIS}\footnote{\url{https://github.com/OpenGVLab/LORIS}}, \textbf{MuVi}~\cite{MuVi}\footnote{\url{https://muvi-v2m.github.io}}, and \textbf{V2Meow}~\cite{V2Meow}\footnote{ \url{https://tinyurl.com/v2meow}}, which introduce innovative frameworks such as Transformer-based, diffusion-based, multi-stage autoregressive models, the absence of publicly available code or pre-trained weights precluded their inclusion in our experiments.

\section{Human Evaluation Details}

\begin{figure}[h]
    \centering
    \includegraphics[width=1\linewidth]{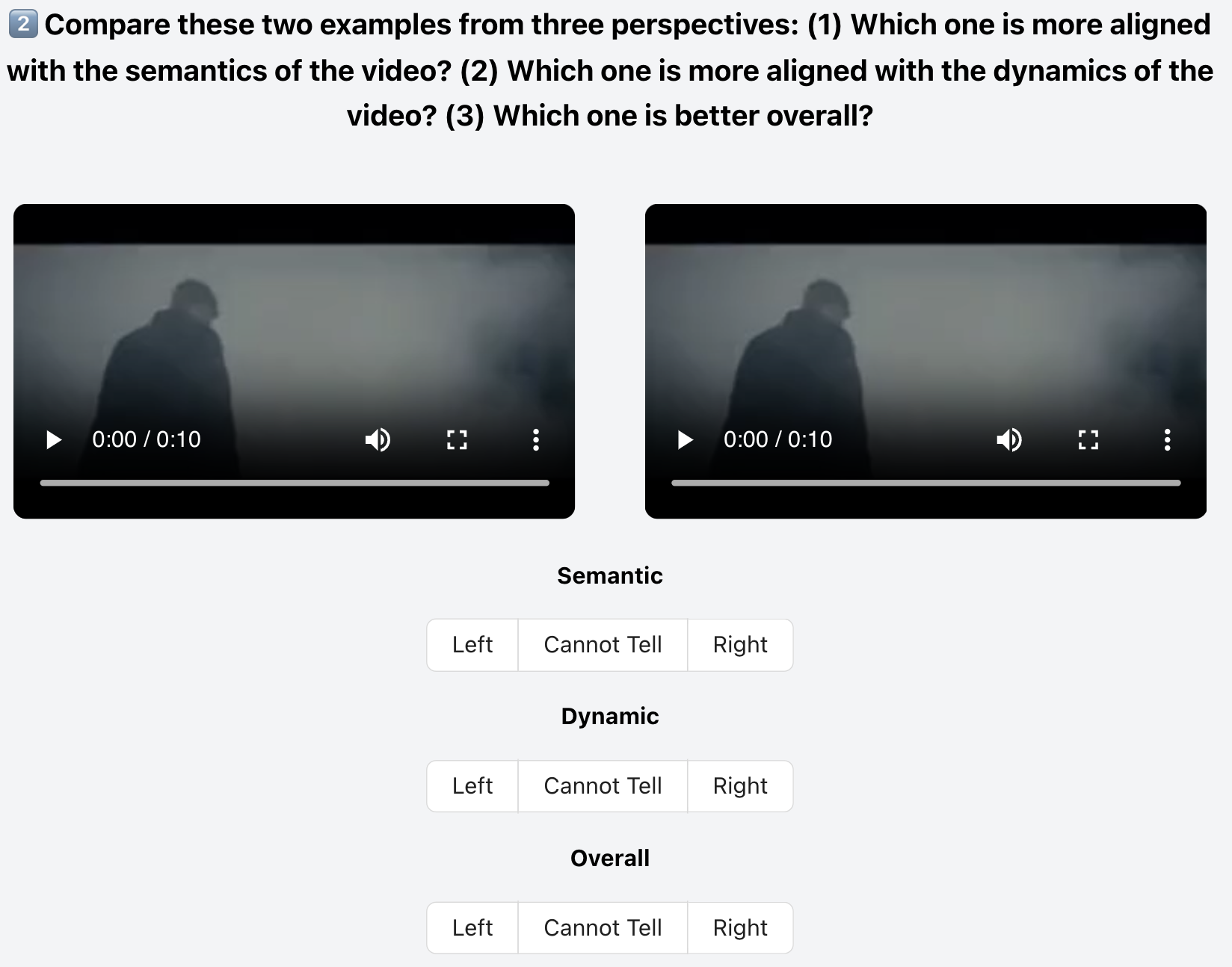}
    \caption{The human evaluation platform for comparing two generated samples. Participants can choose from three options (\ie, left, cannot tell, and right) to express their evaluations.}
    \label{fig:human_compare}
\end{figure}
\begin{figure}[h]
    \centering
    \includegraphics[width=\linewidth]{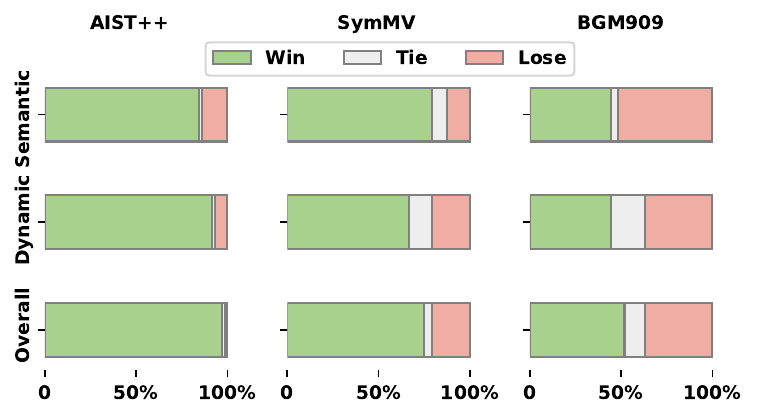}
    \caption{Human voting by comparing music qualities between DyViM (green) and baseline VidMuse (red) from three perspectives: semantic alignment, dynamic alignment, and overall quality.}
    \label{fig:bar}
\end{figure}

To conduct the subjective evaluation for video-to-music generation, we designed a questionnaire that asks raters: ``Rate the generated music from two perspectives: 1) Overall quality (OVL) and 2) Visual relevance (REL).''
Notably, we randomly selected samples for raters to evaluate, collected their ratings, and then averaged the ratings at the end. 
Moreover, we compared DyViM with VidMuse (which is trained on extensive video-music pairs) by asking participants: ``Compare these two examples from three perspectives: 
(1) Which one is more aligned with the semantics of the video?  
(2) Which one is more aligned with the dynamics of the video?  
(3) Which one is better overall?'' This evaluation was conducted on the platform shown in Figure~\ref{fig:human_compare}.
The results are presented in Figure~\ref{fig:bar}. In general, DyViM outperforms VidMuse, showcasing significant improvements, especially on the AIST++ and SymMV datasets.

\newpage

\begin{figure}
    \centering
    \includegraphics[width=\linewidth]{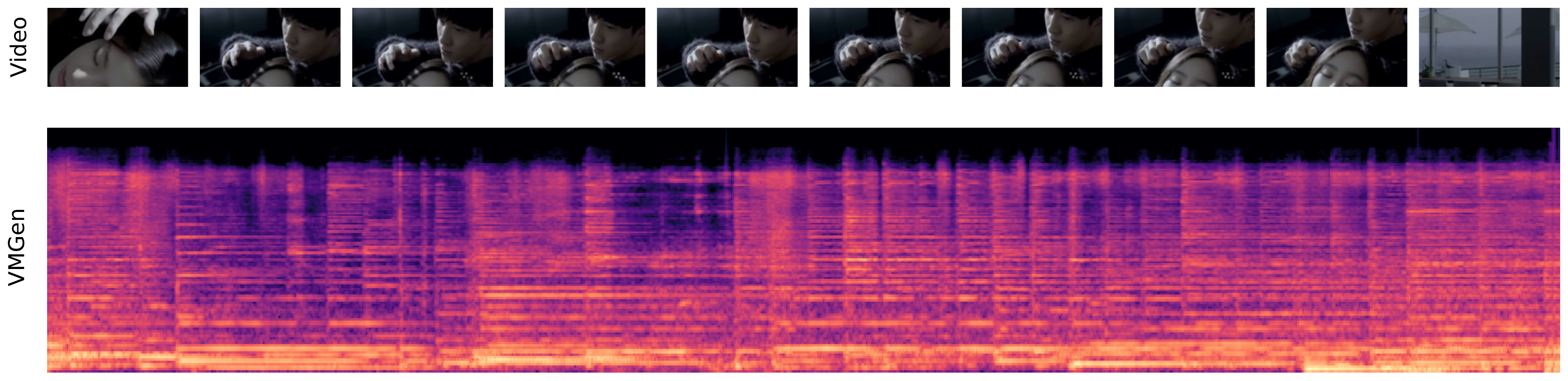}
\end{figure}

\begin{figure}
    \centering
    \includegraphics[width=\linewidth]{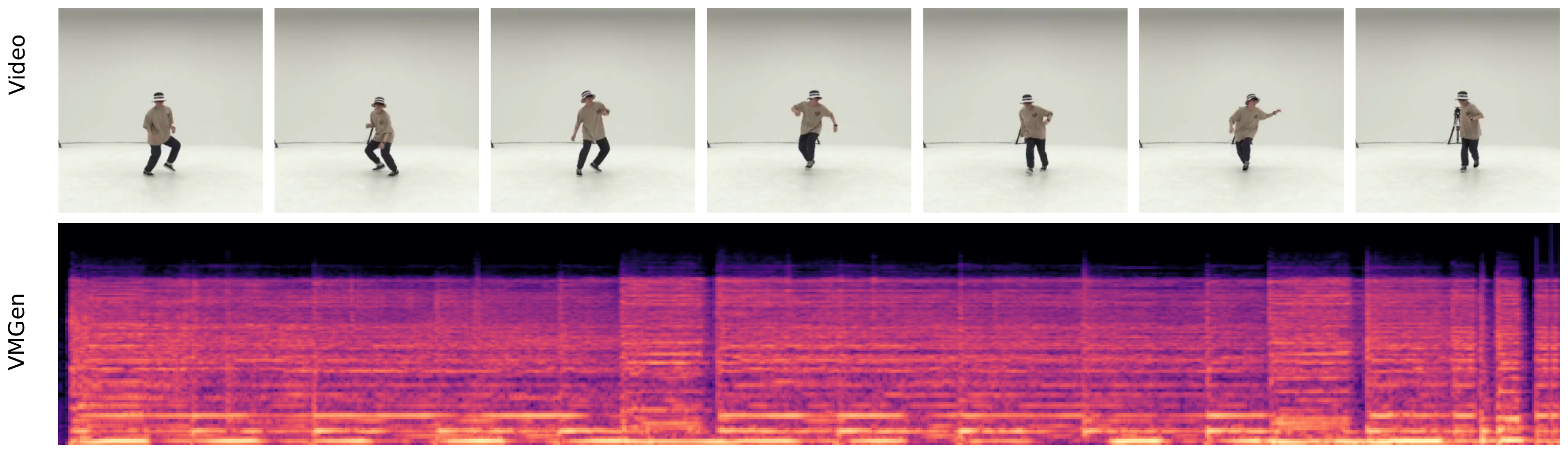}
\end{figure}

\begin{figure}
    \centering
    \includegraphics[width=\linewidth]{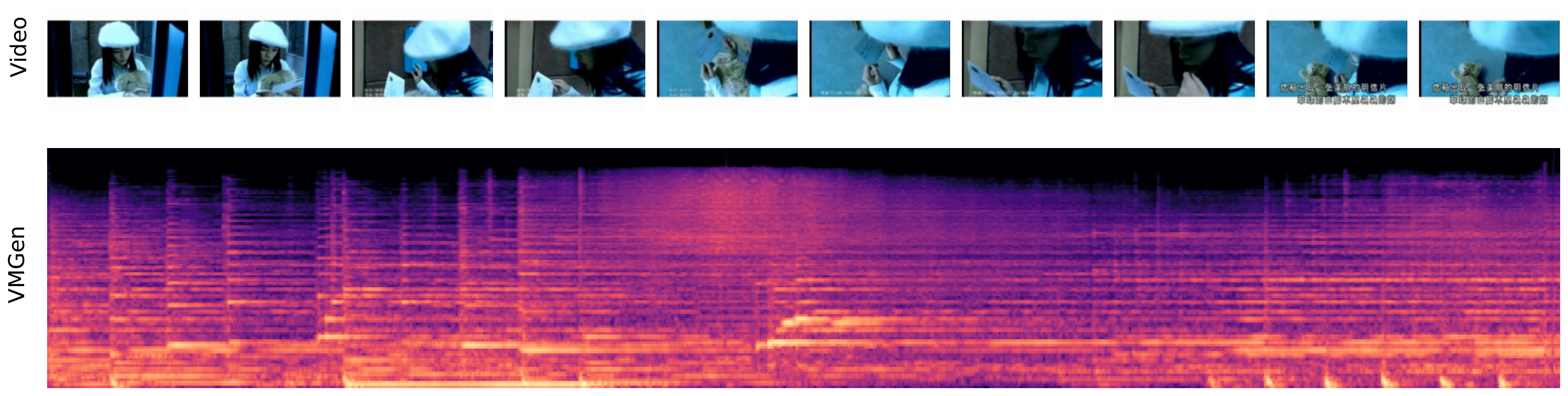}
\end{figure}

\begin{figure}
    \centering
    \includegraphics[width=\linewidth]{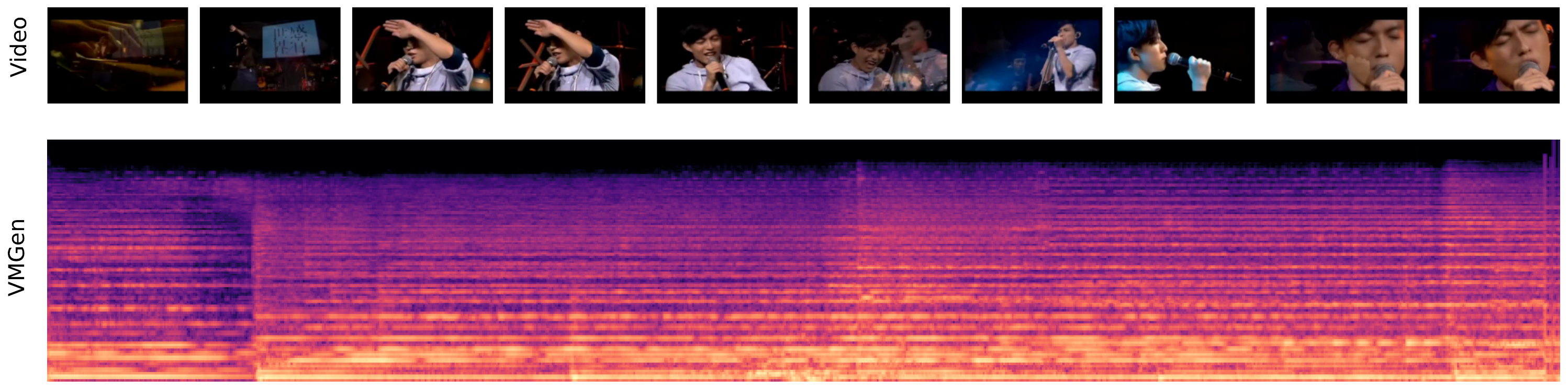}
\end{figure}

\begin{figure}
    \centering
    \includegraphics[width=\linewidth]{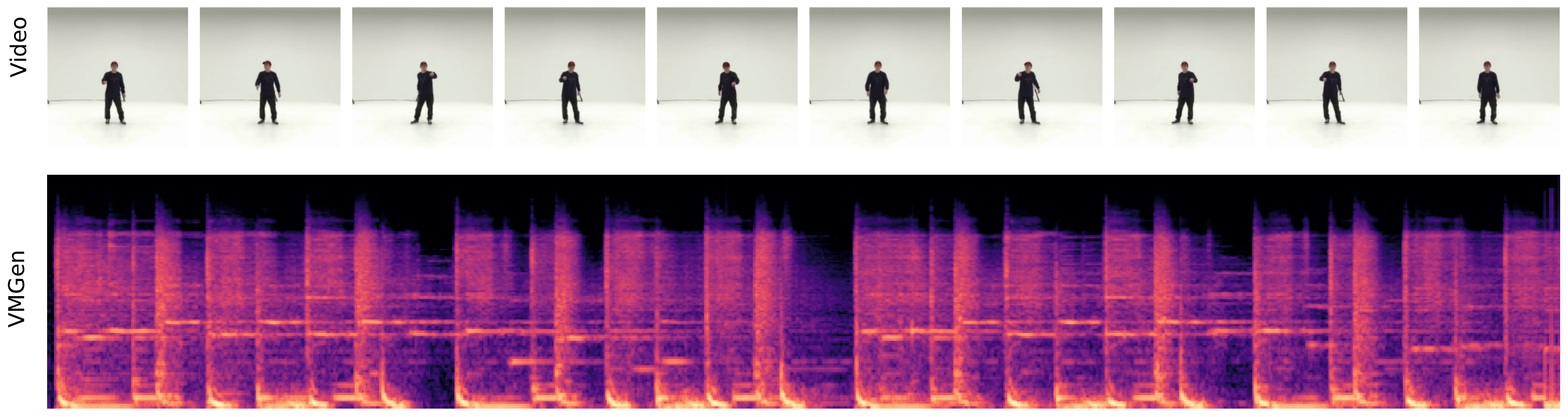}
\end{figure}

\begin{figure}
    \centering
    \includegraphics[width=\linewidth]{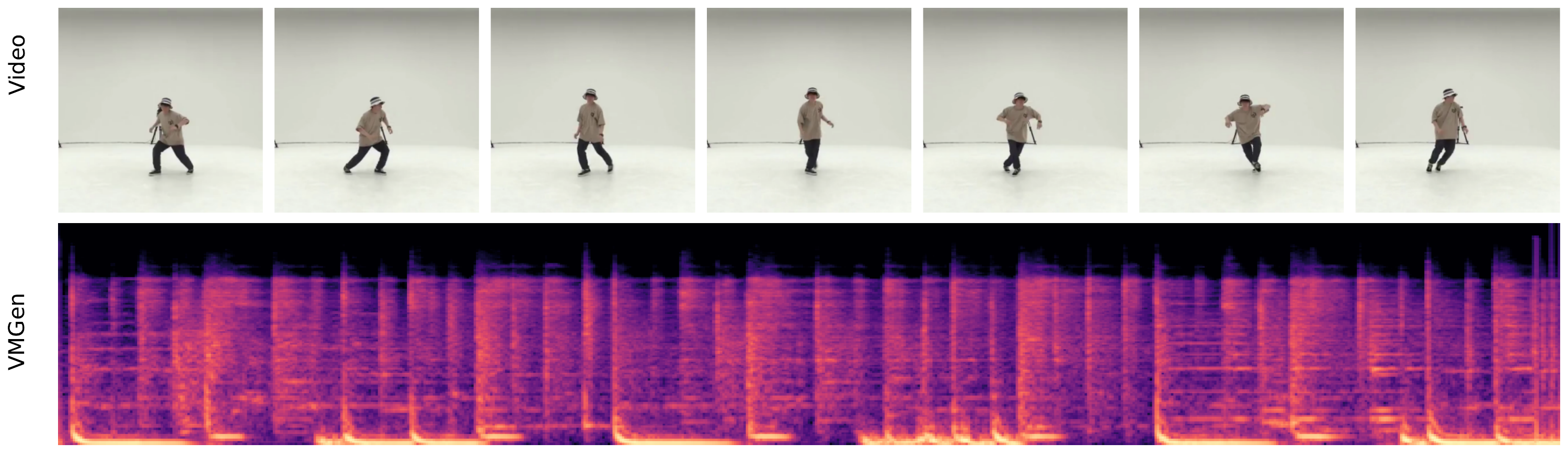}
\end{figure}

\begin{figure}
    \centering
    \includegraphics[width=\linewidth]{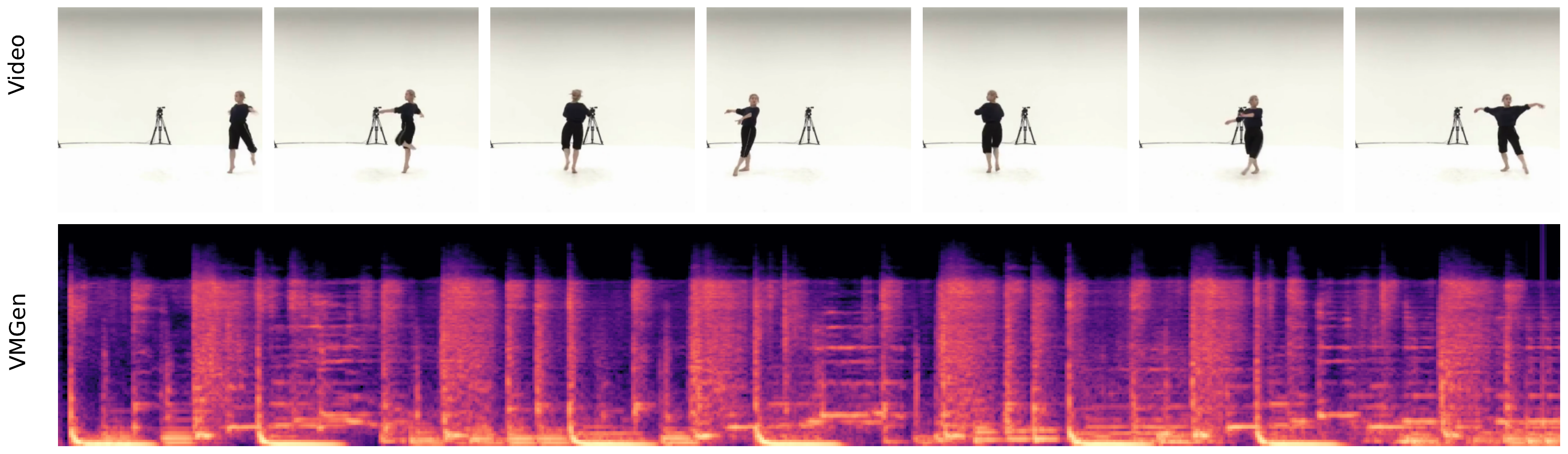}
\end{figure}

\begin{figure}
    \centering
    \includegraphics[width=\linewidth]{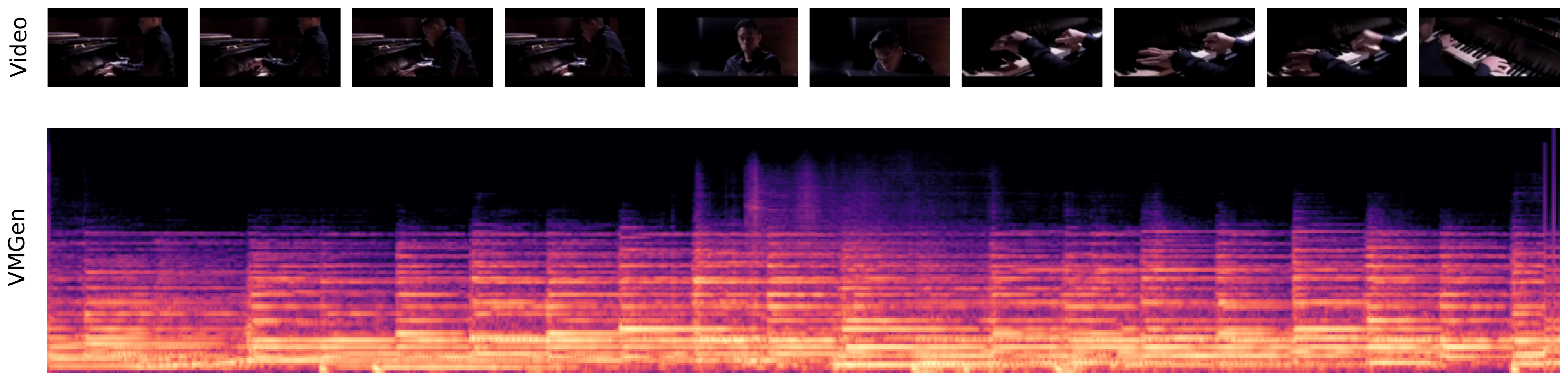}
\end{figure}

\begin{figure}
    \centering
    \includegraphics[width=\linewidth]{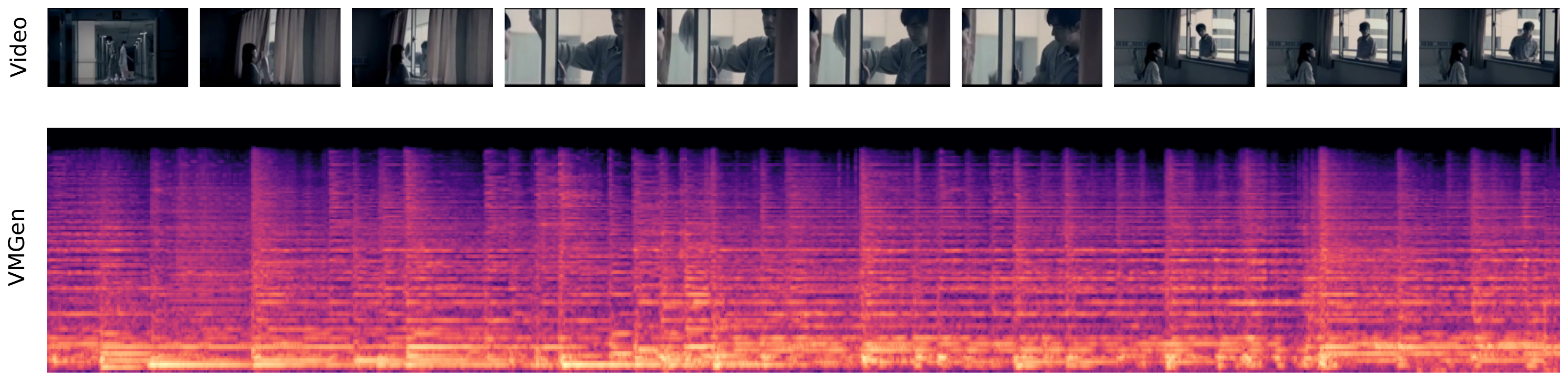}
\end{figure}

\begin{figure}
    \centering
    \includegraphics[width=\linewidth]{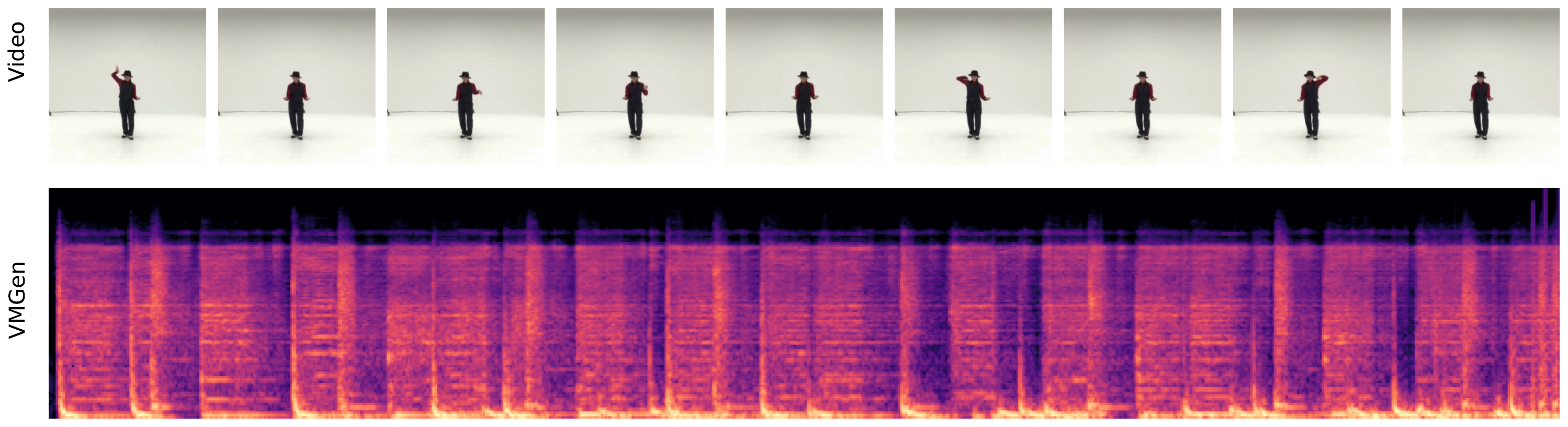}
\end{figure}

\begin{figure}
    \centering
    \includegraphics[width=\linewidth]{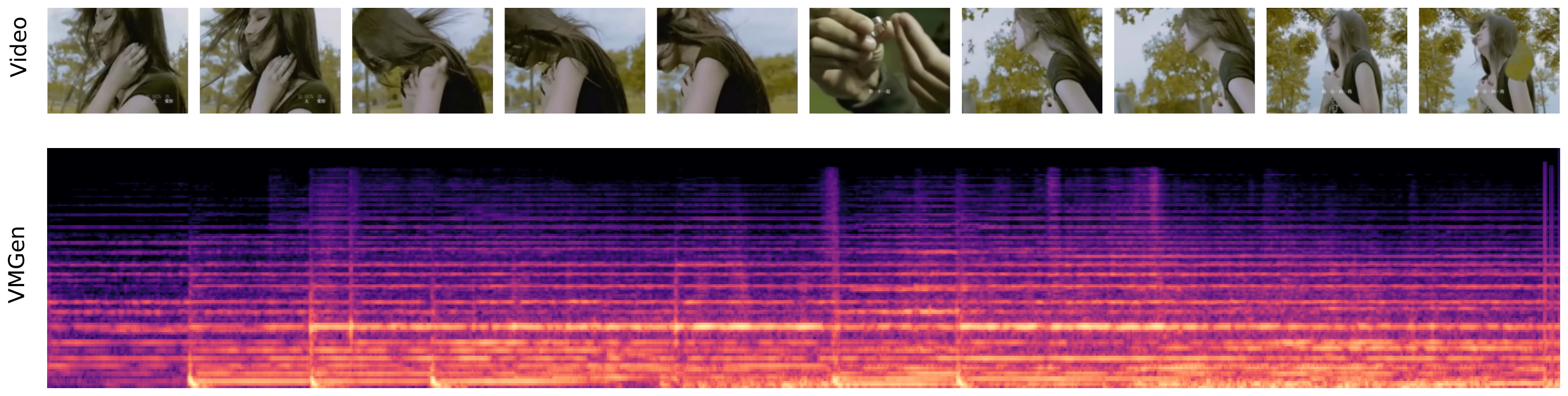}
\end{figure}

\begin{figure}
    \centering
    \includegraphics[width=\linewidth]{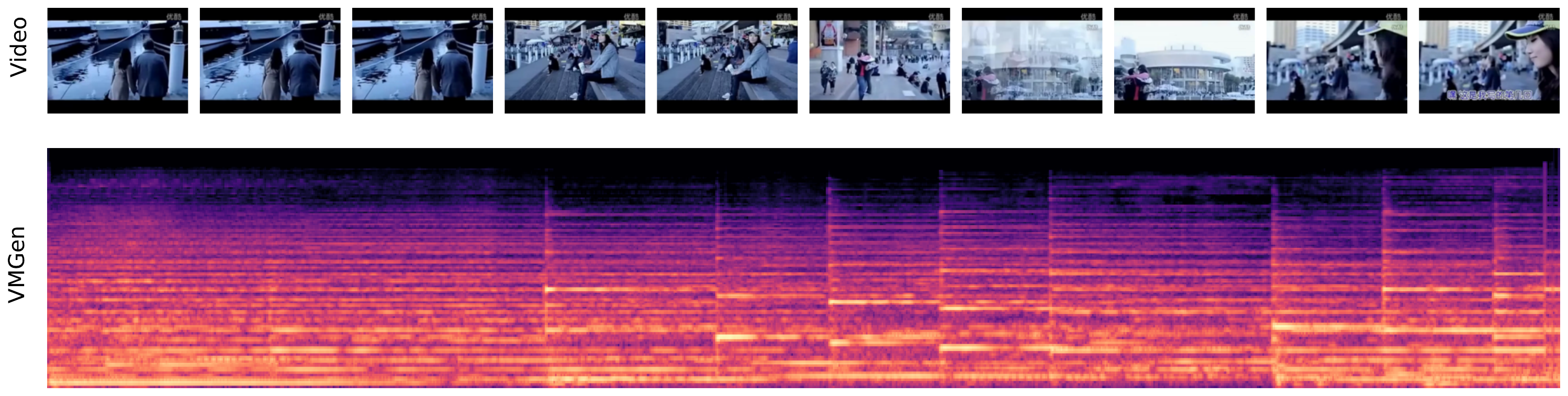}
\end{figure}

\begin{figure}
    \centering
    \includegraphics[width=\linewidth]{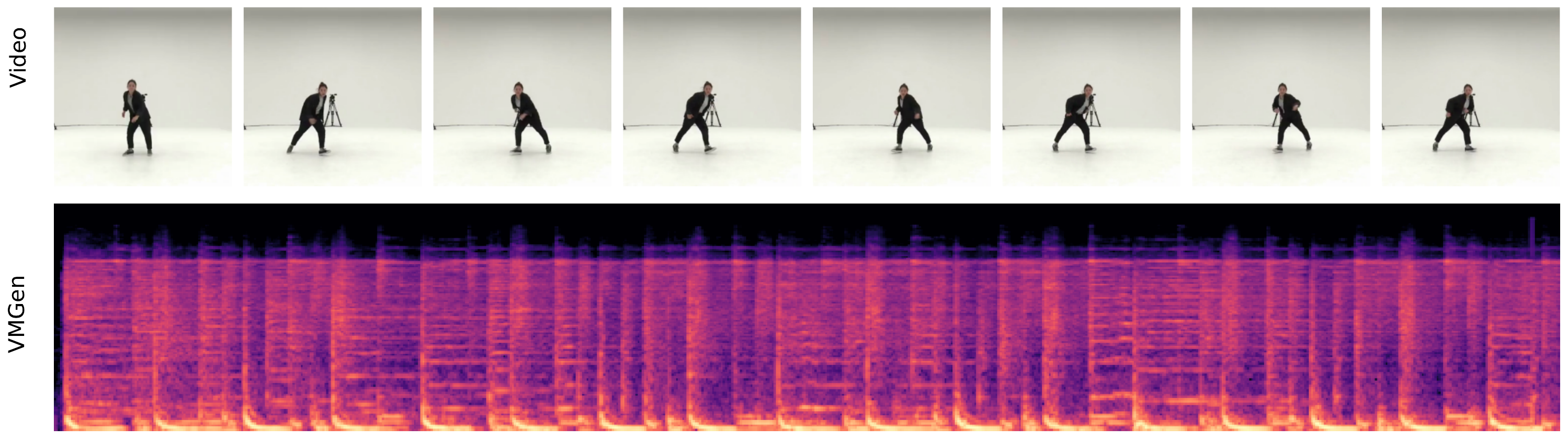}
\end{figure}

\begin{figure}
    \centering
    \includegraphics[width=\linewidth]{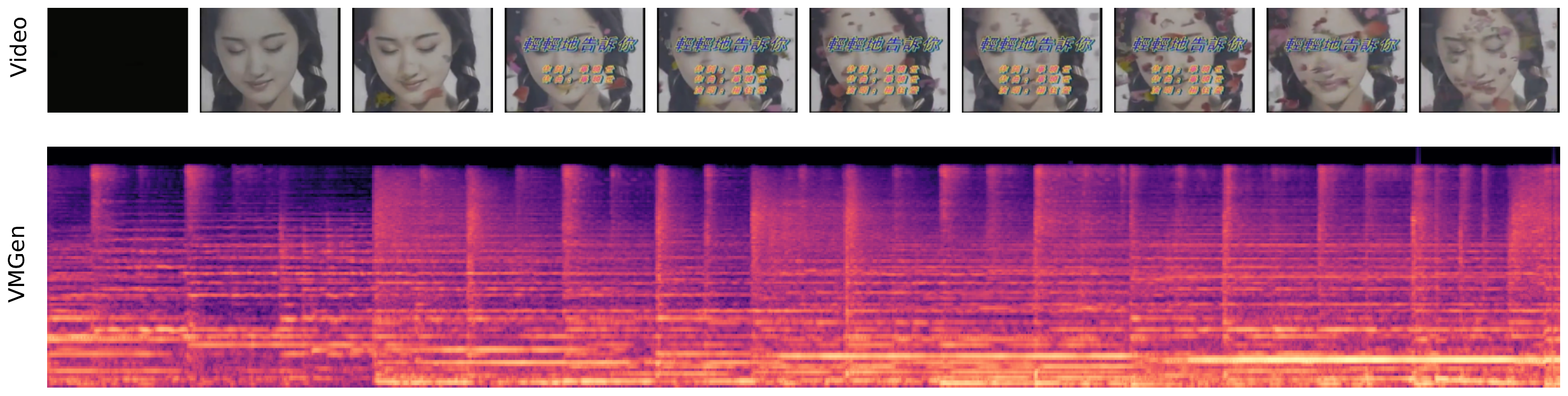}
\end{figure}

\begin{figure}
    \centering
    \includegraphics[width=\linewidth]{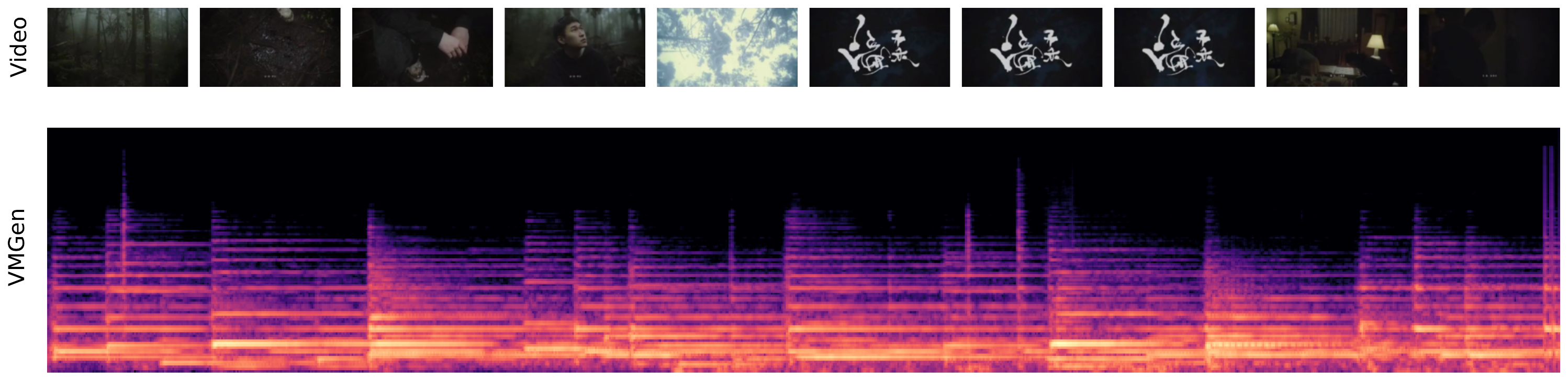}
\end{figure}

\begin{figure}
    \centering
    \includegraphics[width=\linewidth]{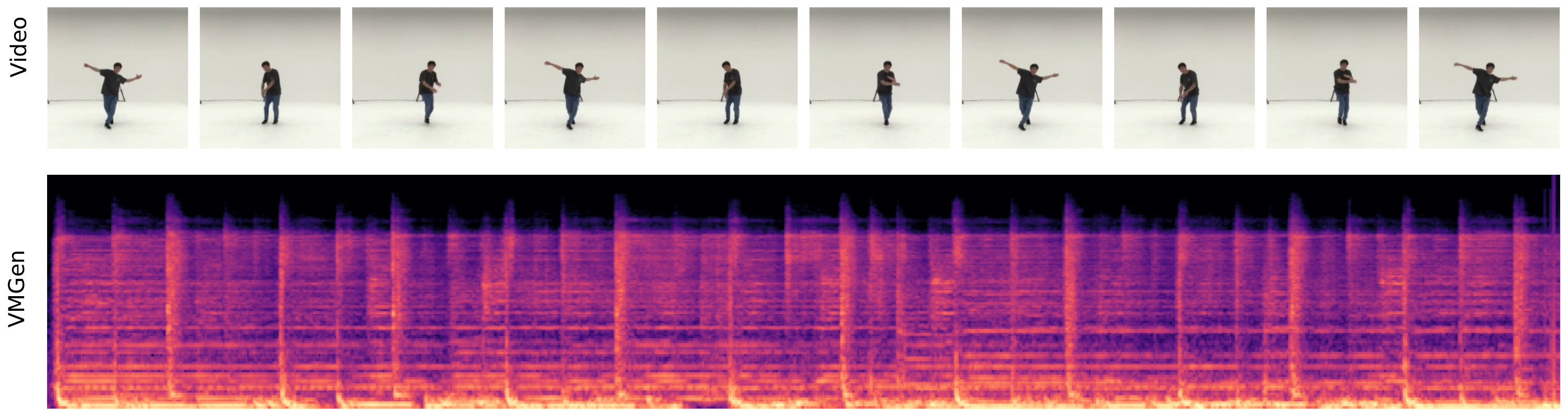}
\end{figure}

\begin{figure}
    \centering
    \includegraphics[width=\linewidth]{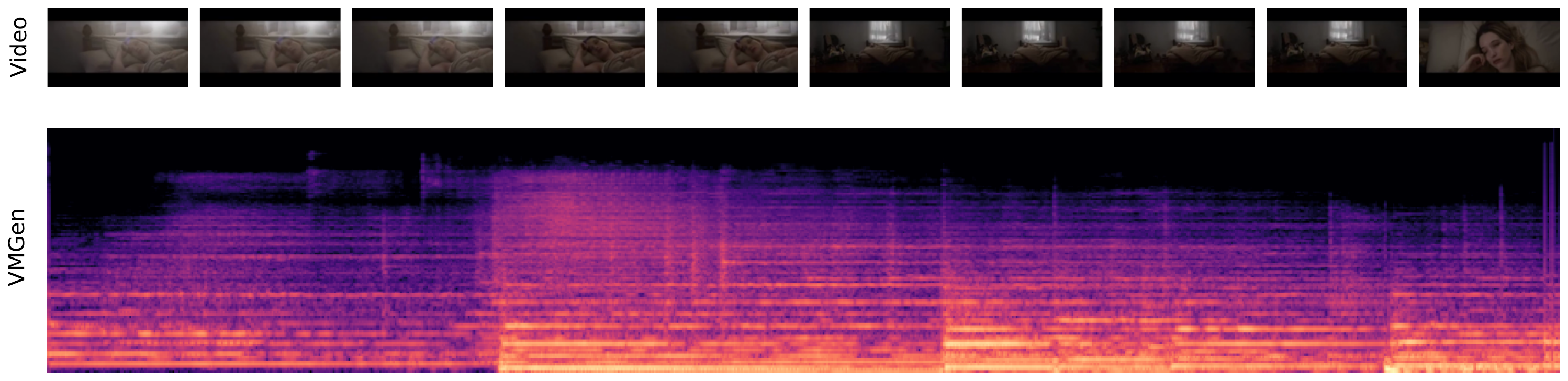}
\end{figure}

\begin{figure}
    \centering
    \includegraphics[width=\linewidth]{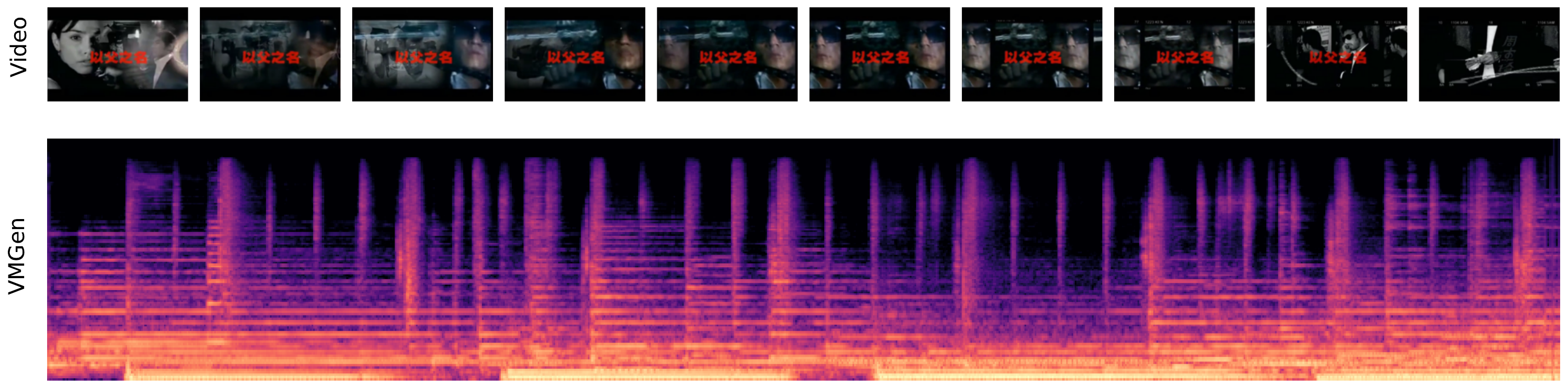}
\end{figure}

\begin{figure}
    \centering
    \includegraphics[width=\linewidth]{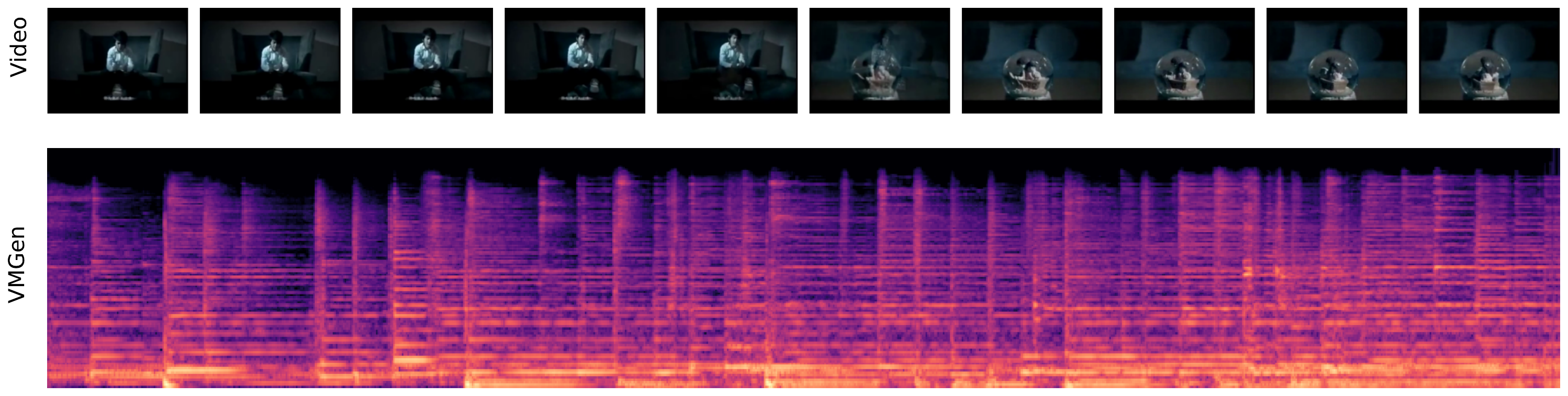}
\end{figure}

\begin{figure}
    \centering
    \includegraphics[width=\linewidth]{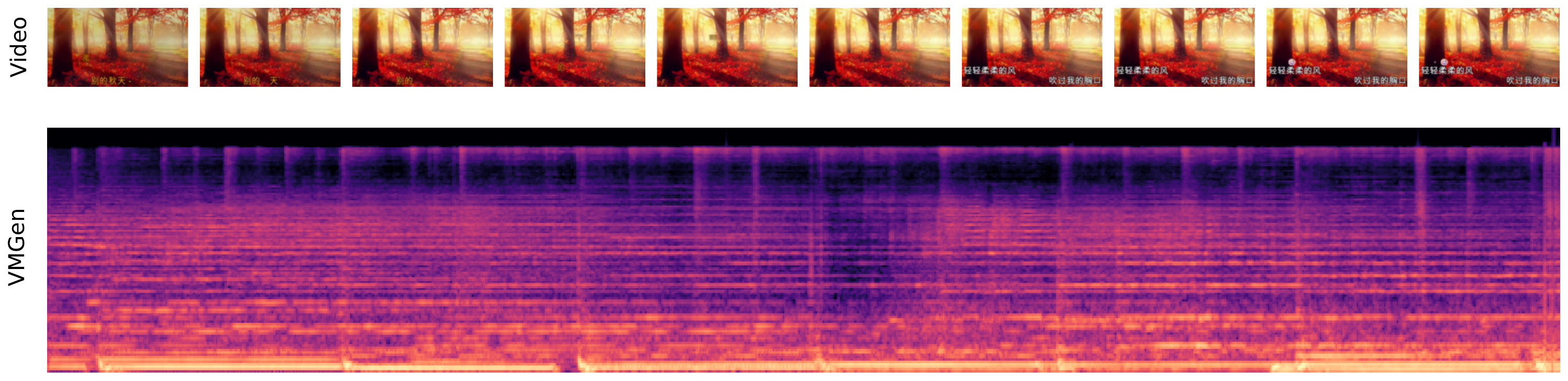}
\end{figure}

\begin{figure}
    \centering
    \includegraphics[width=\linewidth]{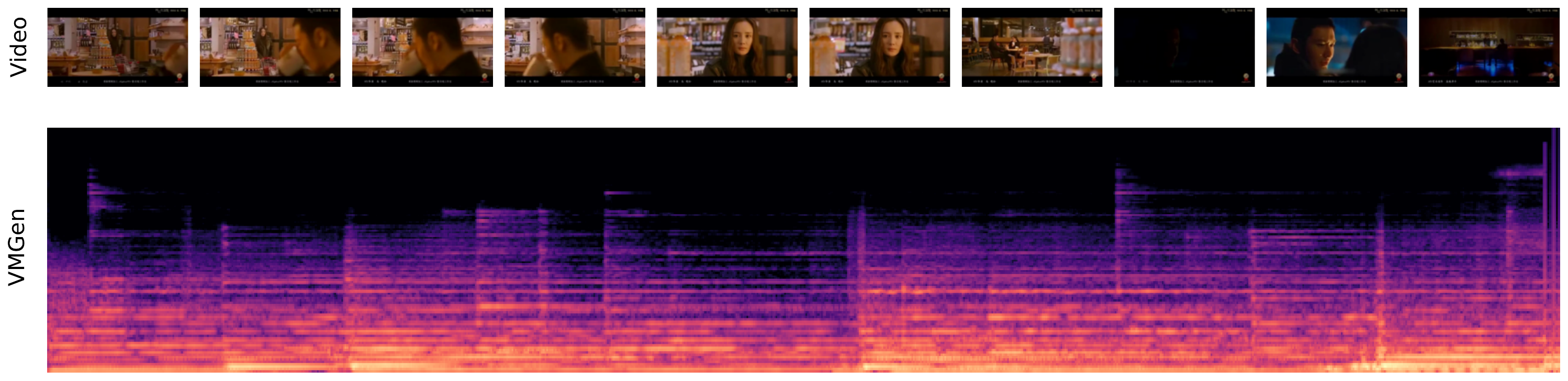}
\end{figure}

\begin{figure}
    \centering
    \includegraphics[width=\linewidth]{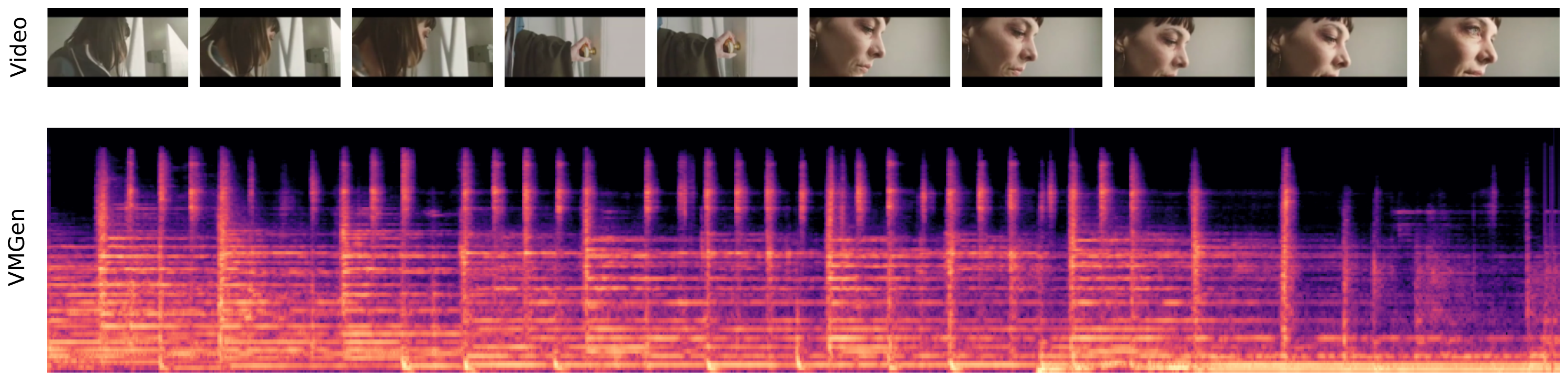}
\end{figure}

\begin{figure}
    \centering
    \includegraphics[width=\linewidth]{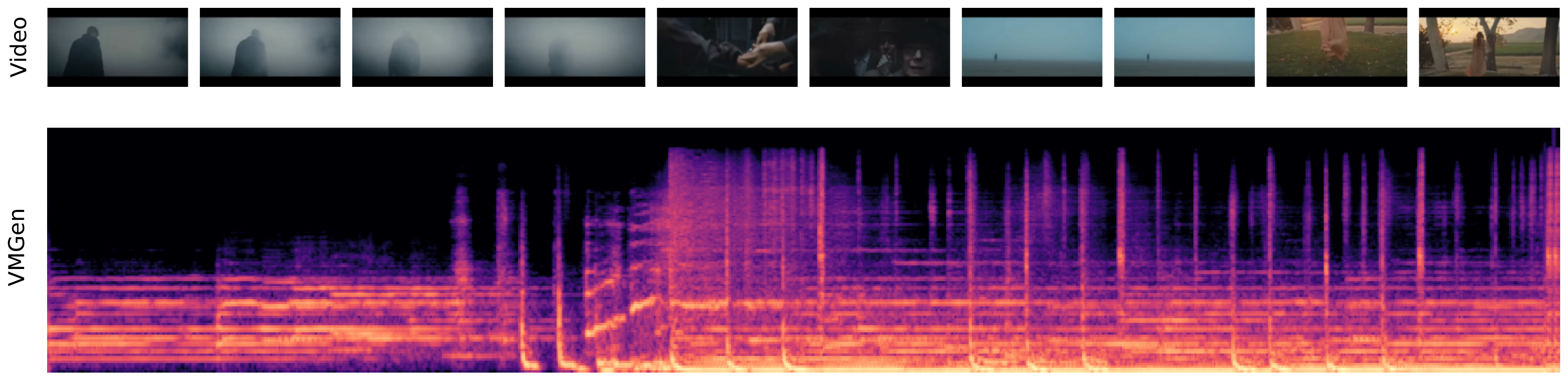}
\end{figure}

\begin{figure}
    \centering
    \includegraphics[width=\linewidth]{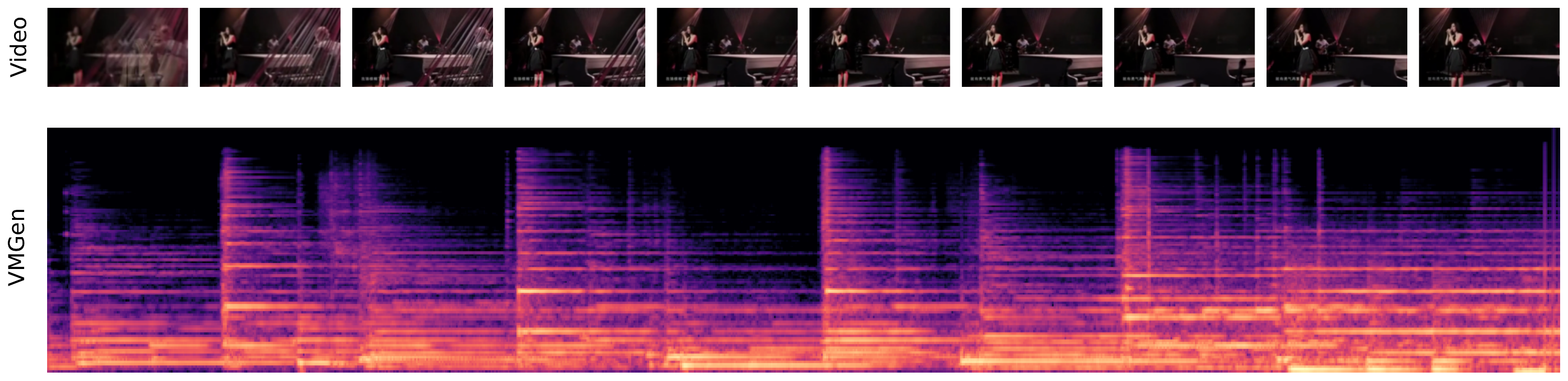}
\end{figure}

\begin{figure}
    \centering
    \includegraphics[width=\linewidth]{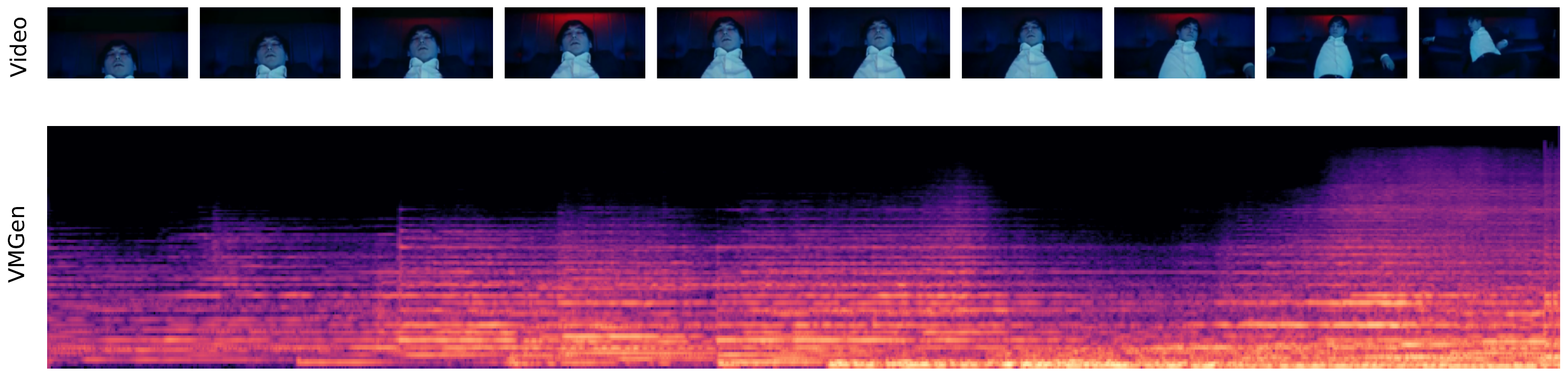}
\end{figure}

\begin{figure}
    \centering
    \includegraphics[width=\linewidth]{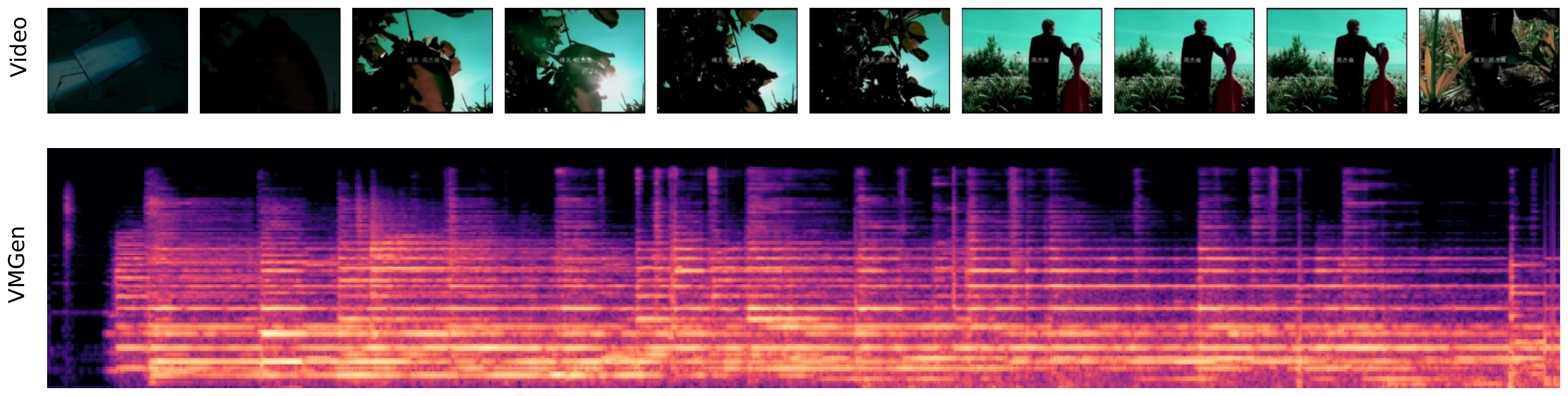}
\end{figure}

\begin{figure}
    \centering
    \includegraphics[width=\linewidth]{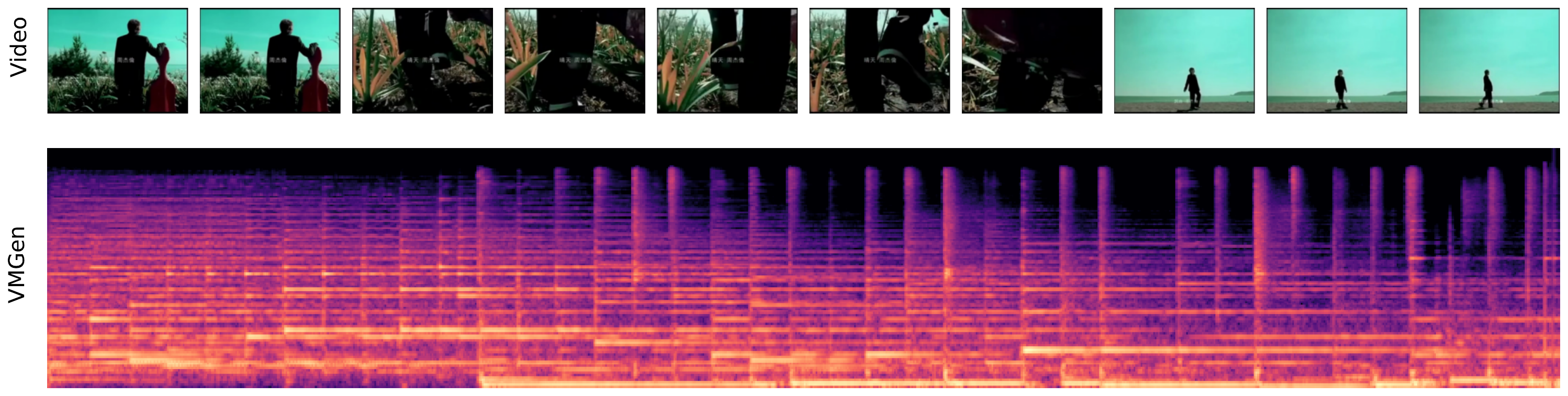}
\end{figure}

\begin{figure}
    \centering
    \includegraphics[width=\linewidth]{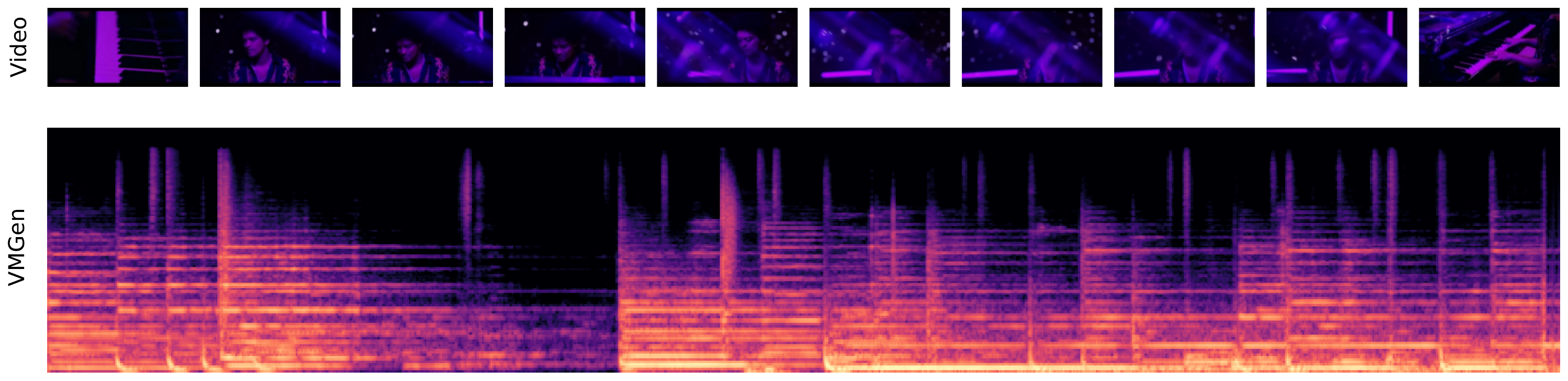}
\end{figure}

\begin{figure}
    \centering
    \includegraphics[width=\linewidth]{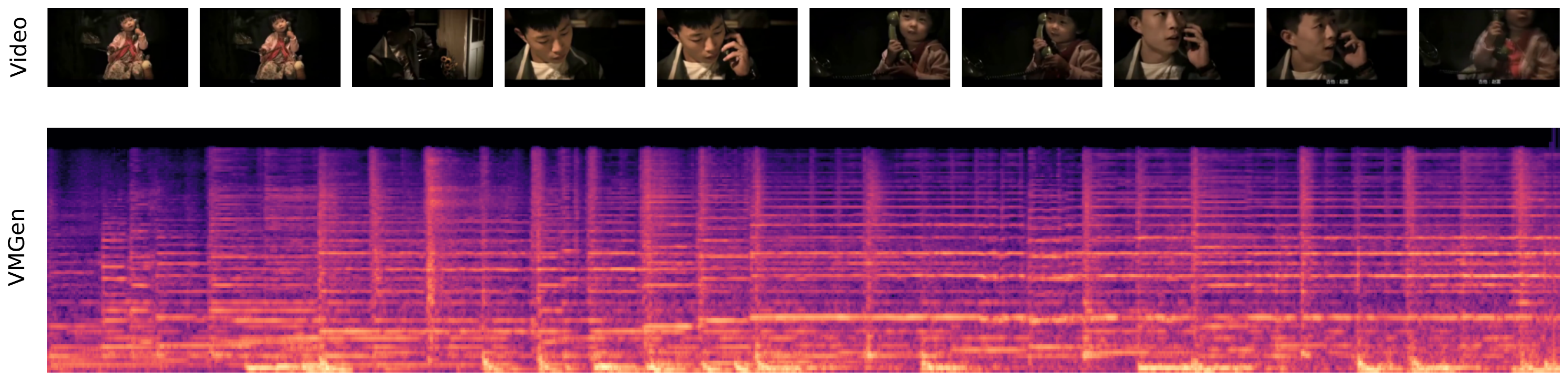}
\end{figure}

\begin{figure}
    \centering
    \includegraphics[width=\linewidth]{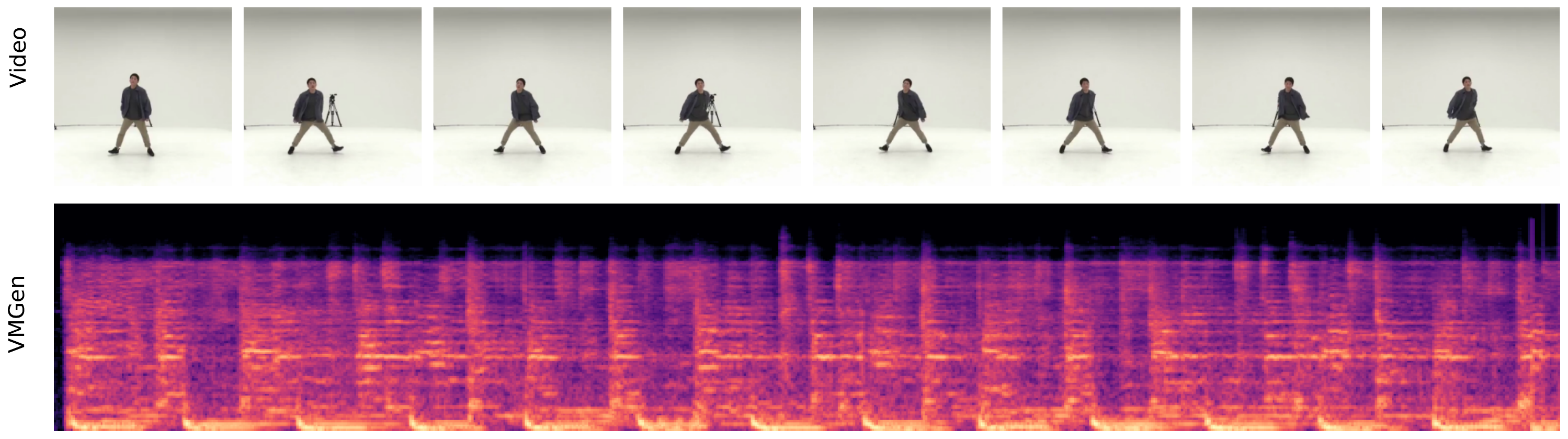}
\end{figure}

\begin{figure}
    \centering
    \includegraphics[width=\linewidth]{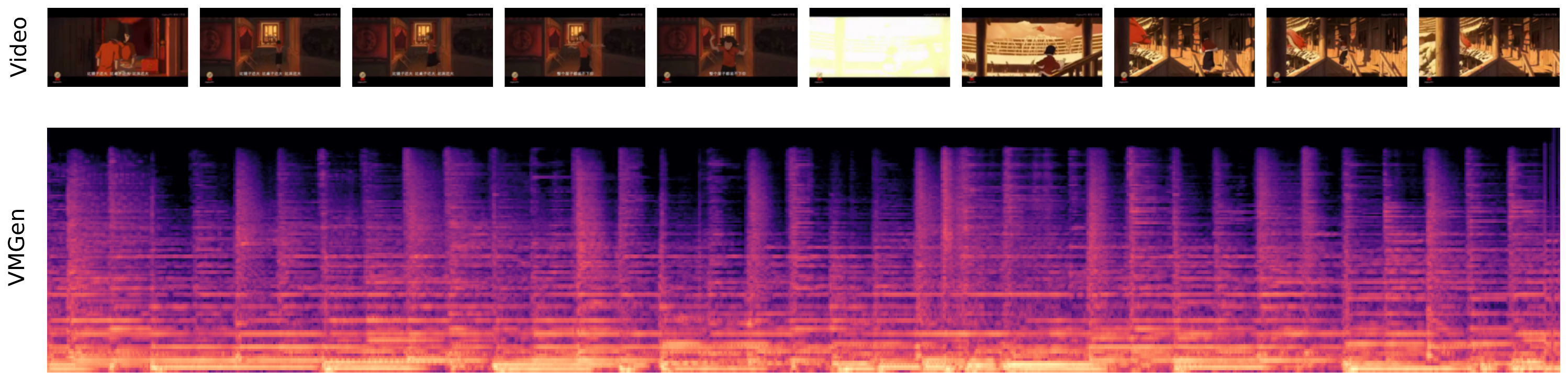}
\end{figure}

\begin{figure}
    \centering
    \includegraphics[width=\linewidth]{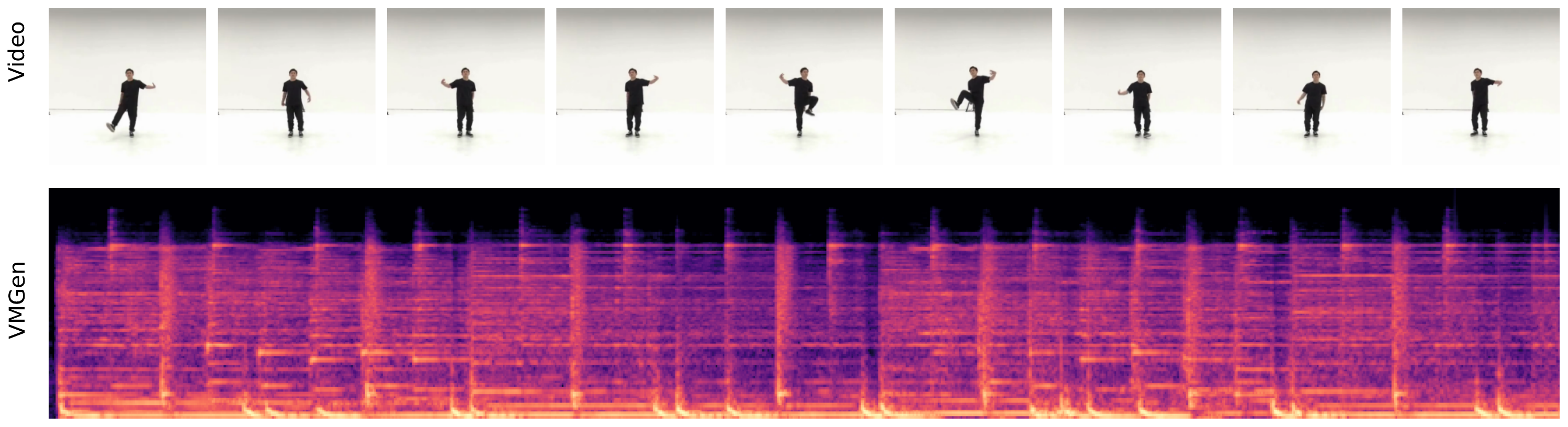}
\end{figure}

\begin{figure}
    \centering
    \includegraphics[width=\linewidth]{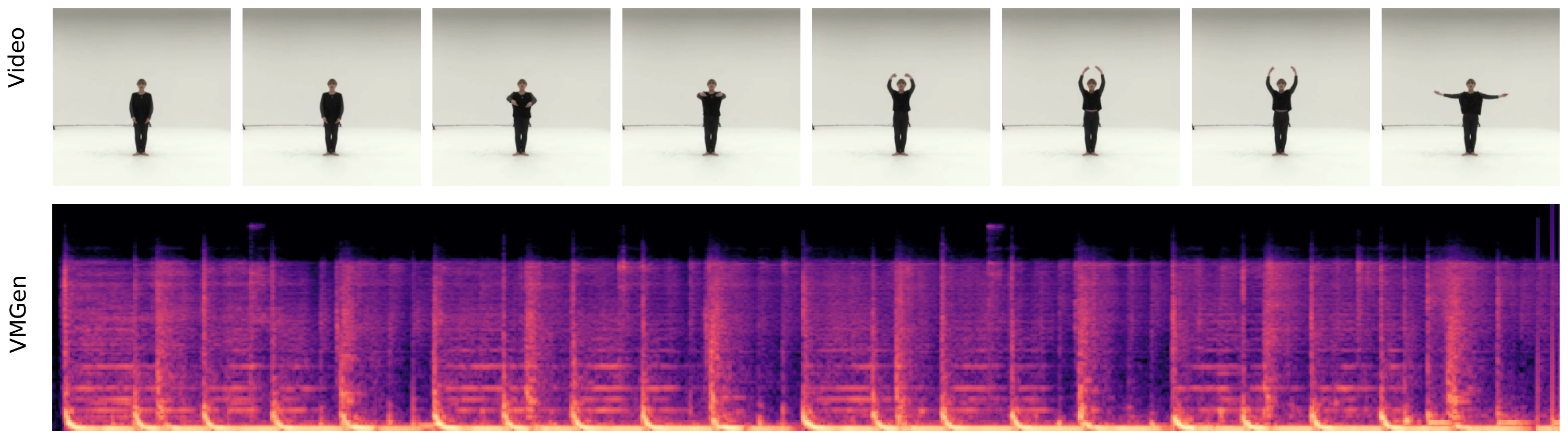}
\end{figure}

\begin{figure}
    \centering
    \includegraphics[width=\linewidth]{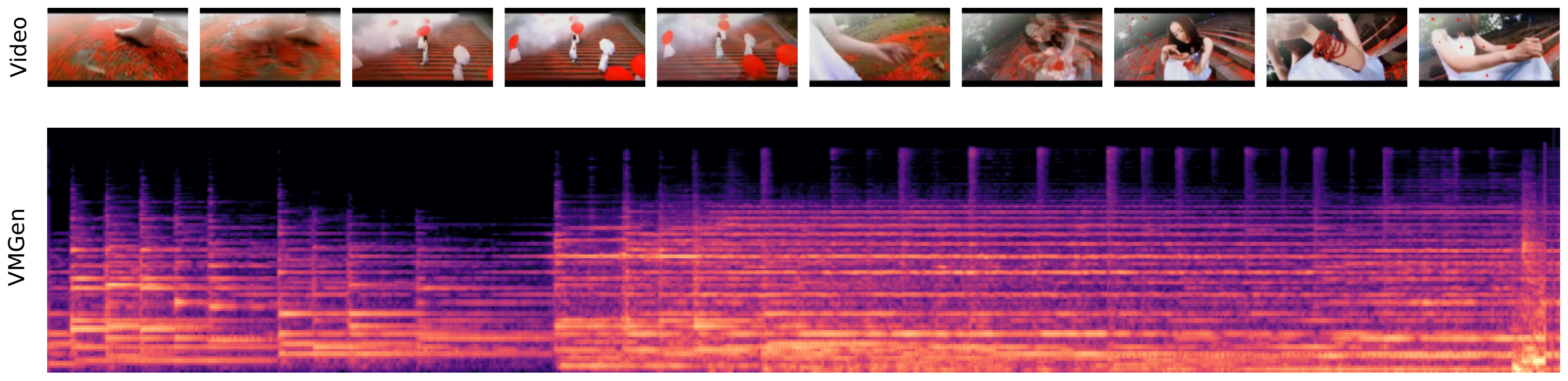}
\end{figure}

\begin{figure}
    \centering
    \includegraphics[width=\linewidth]{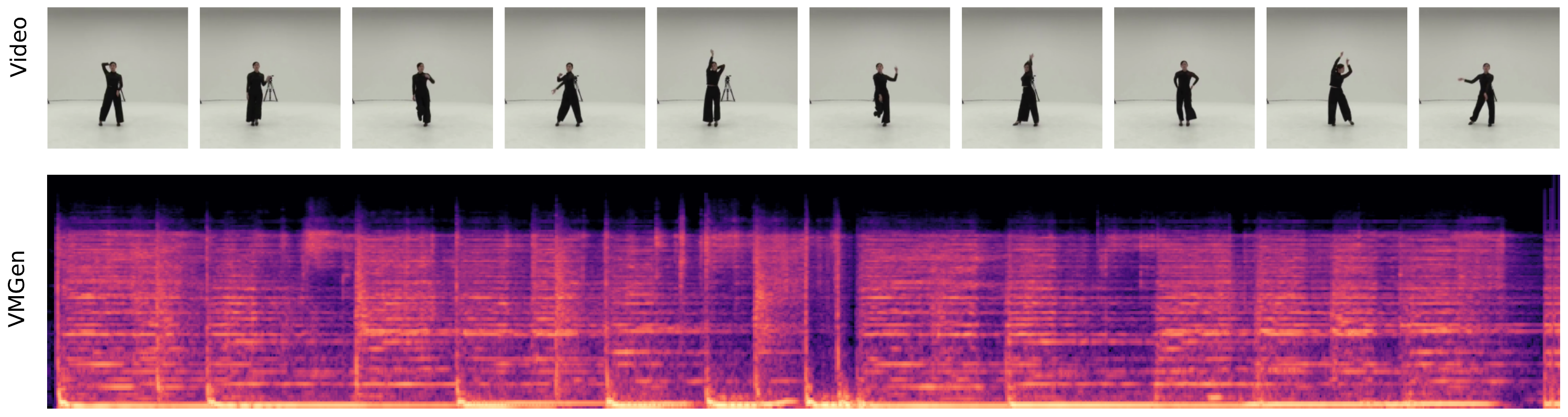}
\end{figure}

\begin{figure}
    \centering
    \includegraphics[width=\linewidth]{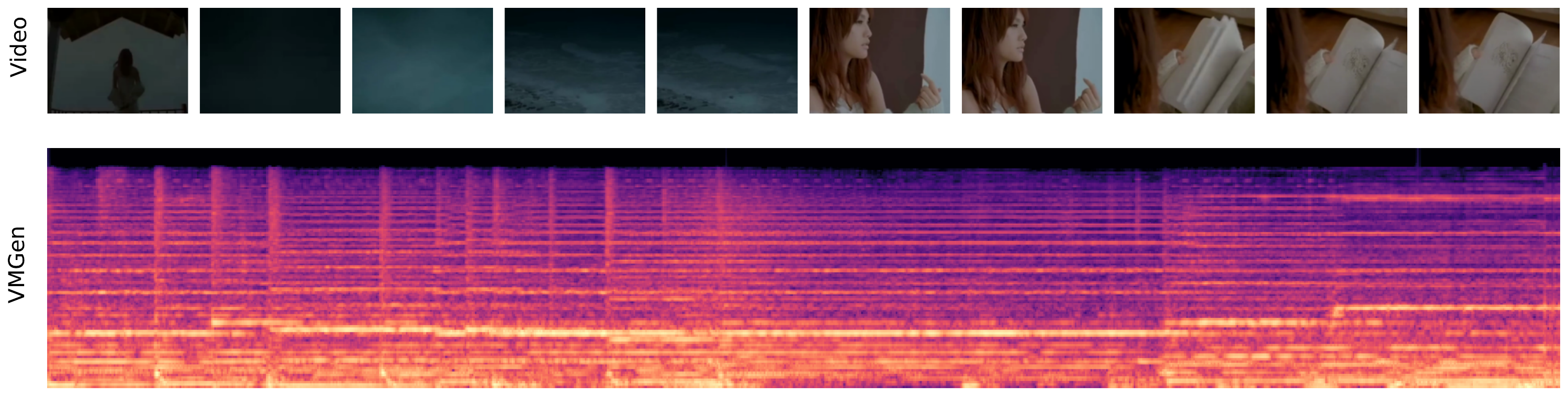}
\end{figure}

\begin{figure}
    \centering
    \includegraphics[width=\linewidth]{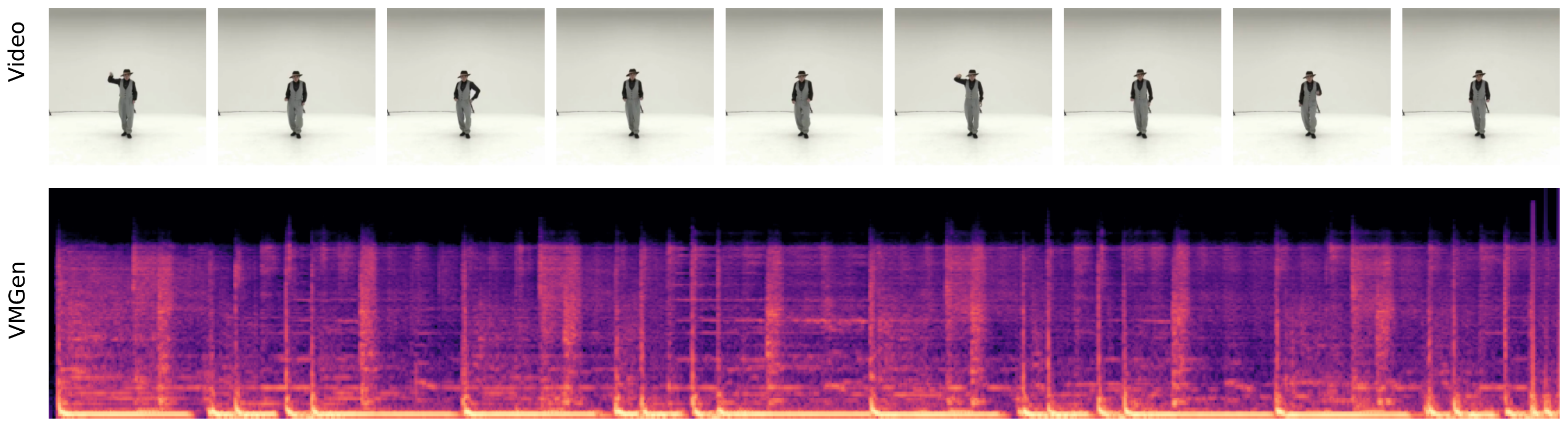}
\end{figure}

\begin{figure}
    \centering
    \includegraphics[width=\linewidth]{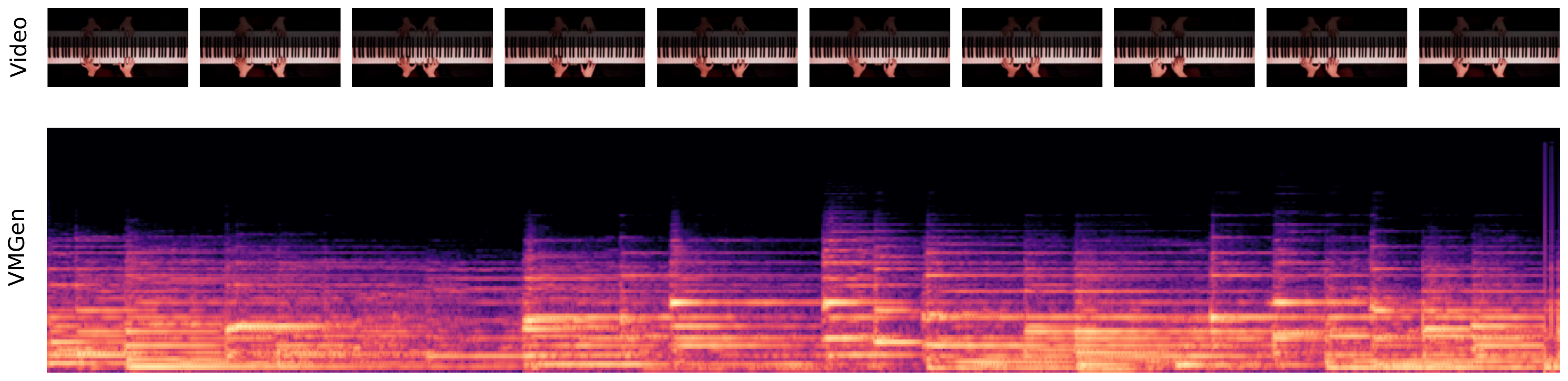}
\end{figure}

\begin{figure}
    \centering
    \includegraphics[width=\linewidth]{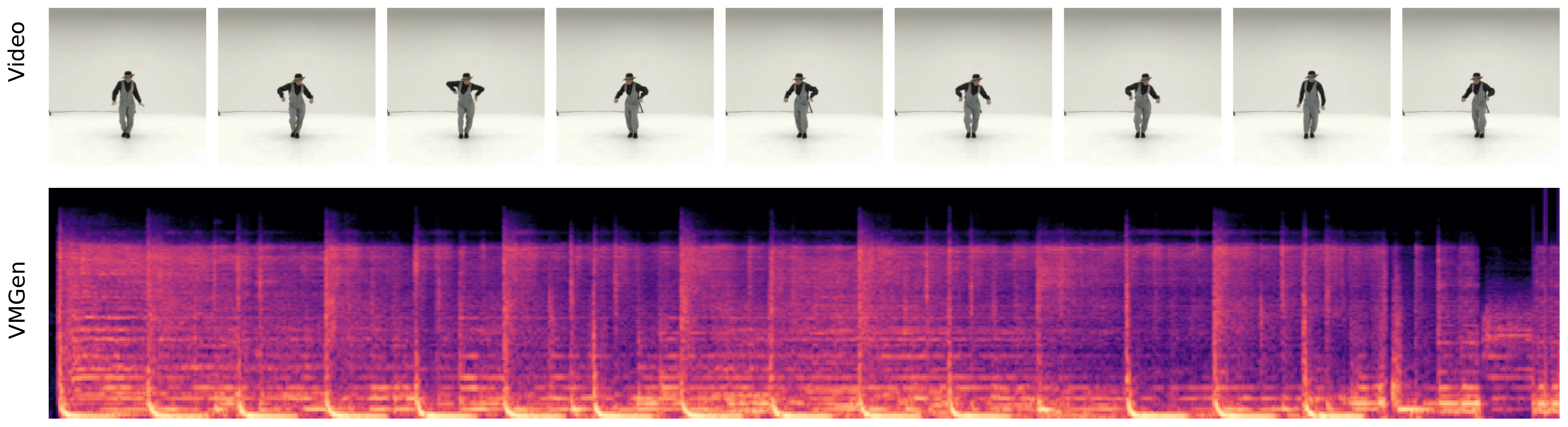}
\end{figure}

\end{document}